\documentclass[12pt]{iopart}

\usepackage[british]{babel}
\usepackage{iopams}
\usepackage{graphicx}
\usepackage{color}

    \definecolor{darkgreen}{rgb}{0.1,0.75,0.1}

    \newcommand{\tconf}{t}

    \newcommand{\be}{\begin{equation}}
    \newcommand{\ee}{\end{equation}}
    \newcommand{\bea}{\begin{eqnarray}}
    \newcommand{\eea}{\end{eqnarray}}
    \newcommand{\bdm}{\begin{displaymath}}
    \newcommand{\edm}{\end{displaymath}}
    \newcommand{\ba}{\begin{array}}
    \newcommand{\ea}{\end{array}}

    \newcommand{\eqref}[1]{Eq.~(\ref{#1})}
    \newcommand{\figref}[1]{Fig.~\ref{#1}}
    
    \newcommand{\secref}[1]{Sec.~\ref{#1}}

    \newcommand{\Mpc}{{\rm Mpc}}
    \newcommand{\Mpci}{{{\rm Mpc}^{-1}}}

    \newcommand{\MeV}{{\rm MeV}}
    \newcommand{\GeV}{{\rm GeV}}
    \newcommand{\Gauss}{{\rm Gauss}}

   \newcommand{\dd}{d}

   \newcommand{\dtconf}{\dd\tconf}

   \newcommand{\dx}{\dd{}x}
   
   \newcommand{\dk}{\dd{}k}
   \newcommand{\dkkk}{\dd^3k}
   \newcommand{\da}{\dd{}a}
   
   \newcommand{\dln}{\dd\ln}
   \newcommand{\dlna}{\dln{}a}
   \newcommand{\dlnk}{\dln{}k}
   
   \newcommand{\order}[1]{{\cal O}(#1)}
   \newcommand{\isorder}[1]{\sim\order{#1}}
\newcommand{\gsim}{\,\raise 0.4ex\hbox{$>$}\kern -0.7em\lower 0.62ex\hbox{$\sim$}\,}

   \newcommand{\V}[1]{{\boldsymbol{#1}}}
   \newcommand{\Vx}{\V{x}}
   \newcommand{\Vy}{\V{y}}
   \newcommand{\Vk}{\V{k}}
   \newcommand{\Vkhat}{\hat{\Vk}}
   \newcommand{\Vq}{\V{q}}

   \newcommand{\VA}{\V{A}}
   \newcommand{\VB}{\V{B}}
   \newcommand{\VE}{\V{E}}
   \newcommand{\VBtilde}{\widetilde{\VB}}
   \newcommand{\VEtilde}{\widetilde{\VE}}
   \newcommand{\Veps}{\boldsymbol{\varepsilon}}

   \renewcommand{\H}{\mathcal{H}}
   \newcommand{\Fdual}{\widetilde{F}}
   \newcommand{\A}{\mathcal{A}}
   \newcommand{\CS}{\mathcal{S}}
   \newcommand{\Rey}{\mathrm{Re}}
   \newcommand{\hel}{\mathfrak{H}}
   \newcommand{\fN}{{f_N}}
   \newcommand{\fNmax}{{f_N^{\max}}}

  \newcommand{\ep}{\epsilon}
  
  \newcommand{\al}{\alpha}
  
  \newcommand{\ra}{\rightarrow}

\begin{document}

% --- TITLE --- %%%%%%%%%%%%%%%%%%%%%%%%%%%%%%%%%%%%%%%%%%%%%%%%%%%%%%%%%

\title[]{Can slow roll inflation induce relevant helical magnetic fields?}

\author[Ruth Durrer, Lukas Hollenstein and Rajeev Kumar Jain]
{Ruth Durrer, Lukas Hollenstein and Rajeev Kumar Jain}

\address{D\'epartement de Physique Th\'eorique, Universit\'e de Gen\`eve, 24,
Quai Ernest Ansermet, CH--1211 Gen\`eve 4, Switzerland}

\ead{
\mailto{Ruth.Durrer@unige.ch},
\mailto{Lukas.Hollenstein@unige.ch},
\mailto{Rajeev.Jain@unige.ch}
}

%%%%%%%%%%%%%%%%%%%%%%%%%%%%%%%%%%%%%%%%%%%%%%%%%%%%%%%%%%%%%%%%%%%%%%%%%
\begin{abstract}
We study the generation of helical magnetic fields during single field
inflation induced by an axial coupling of the electromagnetic field to
the inflaton. During slow roll inflation, we find that such a coupling
always leads to a blue spectrum with $B^2(k) \propto k$, as long as the
theory is treated perturbatively. The magnetic energy density at the end
of inflation is found to be typically too small to backreact on the background
dynamics of the inflaton. We also show that a short deviation from slow
roll does not result in strong modifications to the shape of the spectrum.
We calculate the evolution of the correlation length and the field
amplitude during the inverse cascade and viscous damping of the helical
magnetic field in the radiation era after inflation. We conclude that
except for low scale inflation with very strong coupling, the
magnetic fields generated by such an axial coupling in single field slow
roll inflation with perturbative coupling to the inflaton are 
too weak to provide the seeds for the observed fields in galaxies and
clusters.
\end{abstract}

%%%%%%%%%%%%%%%%%%%%%%%%%%%%%%%%%%%%%%%%%%%%%%%%%%%%%%%%%%%%%%%%%%%%%%%%%

%\pacs{}  % see www.aip.org/pacs
%\maketitle

%%%%%%%%%%%%%%%%%%%%%%%%%%%%%%%%%%%%%%%%%%%%%%%%%%%%%%%%%%%%%%%%%%%%%%%%%
\section{Introduction} \label{sec:intro}
%%%%%%%%%%%%%%%%%%%%%%%%%%%%%%%%%%%%%%%%%%%%%%%%%%%%%%%%%%%%%%%%%%%%%%%%%

Cosmic magnetic fields have been observed on all scales ranging from stars to
near and far away galaxies and galaxy clusters
\cite{Kronberg:1993vk, Pentericci:2000mp, Battaglia:2008ex, Clarke:2000bz}.
The strength of the magnetic fields observed in galaxies and clusters is
typically of the order of $\mu$Gauss. Recently, using the absence of extended
GeV emission around TeV blazar gamma-rays, a lower limit
of $3\times 10^{-16}$ Gauss on the strength of intergalactic magnetic fields
was derived \cite{Neronov:2010,tavecchio-2010.2,tavecchio-2010,Dolag:2010ni}.

These observations prompt the question of the origin of cosmic magnetic fields.
Have they been generated during structure formation or have primordial magnetic
fields been amplified? So far, this question has no clear, satisfactory answer.
Several studies~\cite{Matarrese:2004kq, Gopal:2004ut, Takahashi:2006fea,
Ichiki:2006cd, Kobayashi:2007wd, Ichiki:2007zz, Maeda:2008dv, Fenu:2010kh} of 
second order
perturbation theory have shown that up to recombination only very weak magnetic
fields of the order of $10^{-28}$ Gauss can be generated by structure
formation. However, it turns out that the magnetic fields from second order perturbation
theory are not strong enough to exceed the lower limit derived in
Refs.~\cite{Neronov:2010,tavecchio-2010.2,tavecchio-2010,Dolag:2010ni}.
Whether such fields could have been generated later by the process of galaxy
formation and then ejected into intergalactic space remains unclear.

In this work, we pursue the idea that instead magnetic fields are of primordial
origin and might have been generated in the early universe.
Primordial magnetic fields are interesting as they induce all three kinds
of gravitational perturbations i.e. scalar, vector and tensor; all of which
contribute to the Cosmic Microwave Background (CMB) temperature
and polarisation anisotropies. These primordial fields also lead to non-Gaussian
signals in the CMB even at the lowest order contrary to the higher order effect
due to inflationary scalar perturbations. Analyses of such effects using recent
cosmological data provide an upper limit of a few nGauss on the primordial 
magnetic 
fields~\cite{Durrer:2006pc,Caprini:2009vk,Kahniashvili:2010wm,Trivedi:2010gi}.
It has been argued
that the electroweak or the QCD phase transitions, if they are first order, lead
to the generation of magnetic fields~\cite{Enqvist:1993kf, Joyce:1997uy,
Forbes:2000gr, Boeckel:2009ej}. Also non-perturbative processes during
preheating can generate significant magnetic fields with, in some cases, a
helical component, see for instance \cite{DiazGil:2008tf}. However, causality
strongly constrains such fields. Their power spectrum is very blue with
$B^2(k) \propto k^2$, and therefore, their amplitude on large
scales is far too small~\cite{Durrer:2003ja, Caprini:2001nb}.

On the other hand, if the magnetic fields are produced during inflation, their power
spectrum is a priori not constrained by causality but only by the specific model.
Since the standard electromagnetic (EM) action is conformally invariant the
fluctuations in the EM field are not amplified in the conformally flat expanding
background of inflation. In order to generate magnetic fields, one needs
to break conformal invariance of the EM field, e.g.~by coupling the EM field to a
scalar or a pseudo-scalar field or to a curvature invariant (for an overview see,
for instance, Ref.~\cite{Subramanian:2009fu}). Typically, a term of the form $f F^2$
is considered, where $f$ is a function of time or of the inflaton and $F$ is the EM
field tensor. Depending on the form of the coupling $f$, this gives rise to
different magnetic field power spectra, and even scale-invariant spectra are possible
(in this context, see Refs.~\cite{Martin:2007ue, Demozzi:2009fu}). In this way, the
magnetic fields can have sufficient amplitude on large scales to provide seeds for
the observed fields in galaxies and clusters. However, the backreaction on the
inflation dynamics, the production of gravitational waves and nucleosynthesis bounds
on the amplitude of gravitational waves strongly constrain the magnetic energy
density \cite{Durrer:2006pc, Caprini:2001nb}.

Quite a different situation is encountered if a coupling to the parity-violating
term $F\Fdual$, i.e.~a term $f F\Fdual$, is added to the standard EM action $F^2$,
where $\Fdual$ is the dual of $F$. As a consequence, magnetic helicity is generated,
which is absent in the case discussed above. This has two interesting consequences: firstly,
contrary to non-helical fields, helical fields evolve in the cosmic magnetohydrodynamic
(MHD) plasma via inverse cascade~\cite{Brandenburg:2004jv, Banerjee:2004df, Campanelli:2007tc}.
This transfers power from small to large scales so that even blue spectra can lead to
significant power on large scales. In Ref.~\cite{Caprini:2009pr}, it was shown that the
inverse cascade is not quite sufficient for helical fields generated at the electroweak
phase transition~\cite{Vachaspati:2001nb}, but it might work for magnetic fields from
inflation. Secondly, helical magnetic fields leave a very distinct signature as they
violate parity symmetry. This leads to observable effects, e.g.~correlations between
the anisotropies in the temperature and B-polarisation or in the E- and the
B-polarisations in the CMB~\cite{Caprini:2003vc}.
Furthermore, they induce helical gravitational waves~\cite{Caprini:2003vc} which might
be observable~\cite{Seto:2008sr}.

Some consequences of primordial helical magnetic fields and their generation from
primordial helicity have been studied in the past~\cite{Cornwall:1997ms, Field:1998hi}.
The interactions of helical magnetic fields and axions have also been
investigated~\cite{Campanelli:2005jw, Campanelli:2005ye}. Recently, the generation of
helical magnetic fields during inflation in specific models has been studied,
e.g.~in Ref.~\cite{Anber:2006xt}, helical magnetic fields from N-flation were
investigated. In this case, the large number of pseudo-scalar fields driving inflation
effectively leads to a large coupling, $f\propto\sqrt{N}$, to the $F\Fdual$ term. In
Ref.~\cite{Campanelli:2008kh}, some toy models for the coupling were analysed where
$f$ was taken to be a power law function of $k\tconf$ ($k$ being the wave number and
$\tconf$ the conformal time).

In this paper, we study magnetic fields generated by an axial coupling of the form
$f(\phi)F\Fdual$ during inflation, where $\phi$ is the inflaton. We consider two
different forms of the coupling function and show that, contrary to a non-helical
coupling of the form $f(\phi)F^2$, a helical coupling always leads to a spectral
index $n=1$ for $B^2(k) \propto k^n$, as long as slow roll inflation is considered.
We derive the condition for  the theory to be perturbative, i.e. the free part of the action
dominates over the interaction. Of course this is not necessarily true, but at least naively
if not, we can no longer trust our calculation of particle creation out of the
quantum vacuum which is based on perturbative quantum field theory. We estimate the
magnetic energy density as a function of scale and show that backreaction is
typically small. These conclusions are valid for any reasonable coupling function
$f(\phi)$. We confirm our analytical results numerically. Furthermore, we study the
effects of a short deviation from slow roll on the magnetic field spectrum and show
that such deviations, if kept within the bounds permitted by the CMB data, do not
strongly modify its shape. Even though the inverse cascade in the radiation
dominated era after inflation does move power to larger scales, the final strength
of the magnetic field on cosmologically interesting scales is still 
insufficient to provide seeds for the observed magnetic fields in galaxies and
clusters, except if the inflation scale is low, $T_\ast  < 10^4$GeV and the 
axial coupling is very strong. Even if we assume very efficient dynamo amplification.

The remainder of this paper is organised as follows. In the next section, we
introduce the axial coupling of the EM field to the inflaton, derive the field
equations, discuss the background evolution and the slow roll approximation and
derive the linear perturbation equations of the inflaton and the EM vector potential.
In~\secref{sec:modeeq}, we discuss the evolution equation of EM quantum fluctuations
during inflation and compare the helical to the normal case.
In~\secref{sec:slowroll}, we derive a condition on the chiral coupling by
requiring the theory to be perturbative, investigate different coupling functions in
the slow roll approximation, and solve the evolution of the EM fluctuations
analytically. In~\secref{sec:numerical}, we discuss the consequences of a brief
violation of the slow roll approximation on the magnetic field spectrum.
In~\secref{sec:interpretation}, we study the evolution of the magnetic field during
the radiation era after inflation and determine the final spectrum after the inverse
cascade. Finally, in~\secref{sec:conclusions}, we conclude with a few comments on our
results. Three appendices contain some details on the quantisation of the vector
potential, the condition on the coupling function from the perturbativity of the
theory and the asymptotic behaviour of the Coulomb wave functions, respectively.

%%%%%%%%%%%%%%%%%%%%%%%%%%%%%%%%%%%%%%%%%%%%%%%%%%%%%%%%%%%%%%%%%%%%%%%%%
\paragraph{Notation and units:}
We work in a metric with signature \mbox{($-$ + + +)}. For tensor components, Greek
indices take values $0\ldots3$, while Latin indices run from $1$ to $3$. The components
of spatial $3$-vectors with respect to a comoving basis are denoted in bold face.
We employ Heaviside-Lorentz units such that $c=\hbar=k_B=\epsilon_0=\mu_0=1$.
The reduced Planck mass is defined as $m_P=(8\pi G)^{-1/2}$. We normalise the cosmic
scale factor to unity today so that the comoving scales become physical scales today.

%%%%%%%%%%%%%%%%%%%%%%%%%%%%%%%%%%%%%%%%%%%%%%%%%%%%%%%%%%%%%%%%%%%%%%%%%
\section{Axial coupling of electromagnetism to the inflaton} \label{sec:basics}
%%%%%%%%%%%%%%%%%%%%%%%%%%%%%%%%%%%%%%%%%%%%%%%%%%%%%%%%%%%%%%%%%%%%%%%%%

%%%%%%%%%%%%%%%%%%%%%%%%%%%%%%%%%%%%%%%%%%%%%%%%%%%%%%%%%%%%%%%%%%%%%%%%%
\subsection{Action and field equations}

We consider a scalar field, $\phi$, which takes the role of the inflaton and the EM
field, $F_{\mu\nu}\equiv\partial_{\mu}A_{\nu}-\partial_{\nu}A_{\mu}$, characterised
by its four-vector potential $A_\mu$. The EM field is conformally coupled to the
metric and therefore, no fluctuations are generated unless there is either an
explicit coupling to the inflaton or conformal symmetry is broken directly, e.g.~by
coupling $F$ to a curvature term. Here, we investigate the first possibility and
study a helical coupling given by the action
\be
  S[\phi,A_\mu]\ \equiv\ \int\dd^4x \sqrt{-g} \left\{
    \mathcal{L}_\phi(\phi) + \mathcal{L}_{\rm em}(A_\mu) + \mathcal{L}_{I}(\phi,A_\mu)  \right\}\,.
\ee
The Lagrangian densities of the free fields are
\bea
  \mathcal{L}_\phi(\phi)\ \equiv\ \frac{1}{2}g^{\alpha\beta}(\partial_\alpha\phi)
    (\partial_\beta\phi) + V(\phi)
  \\
  \mathcal{L}_{\rm em}(A_\mu)\ \equiv\ -\frac{1}{4}F_{\alpha\beta}F^{\alpha\beta} \,.
\eea
and the axial interaction is given by
\be
  \mathcal{L}_{I}(\phi,A_\mu)\ \equiv \frac{1}{4} f(\phi) F_{\alpha\beta}\Fdual^{\alpha\beta} \,.
\ee
It describes a coupling of the scalar field to the parity-violating term,
$F\Fdual$, where $\Fdual$ is the dual of the EM field tensor and is defined as
\be \label{e:fdual}
  \Fdual^{\mu\nu}\ \equiv\ \frac{1}{2}\eta^{\mu\nu\alpha\beta}F_{\alpha\beta} \,.
\ee
Here $\eta^{\mu\nu\alpha\beta}$ is the totally anti-symmetric tensor in four
dimensions with $\eta^{0123} \equiv (-g)^{-1/2}$. For an observer with
4-velocity $u^\mu$ the electric and magnetic fields are
$E_\mu=F_{\mu\alpha}u^\alpha$ and $B_\mu=\Fdual_{\mu\alpha}u^\alpha$,
respectively, and we have
$F_{\alpha\beta}\Fdual^{\alpha\beta}=-4E_\alpha B^\alpha$.

If the scale of inflation is above the electroweak scale then, in principle,
one should specify whether $\phi$ couples to $U(1)$, $SU(2)$ or both
of the electroweak $SU(2)_L\times U(1)_Y$. If the coupling is to both,
which seems simplest, the same process that leads to magnetic field helicity
also induces a non-zero baryon number which is related to the electroweak
Chern-Simons number~\cite{Vachaspati:2001nb}. The electromagnetic Chern-Simons 
number is equivalent to the helicity. In this work we do not discuss this 
additional potentially interesting aspect, but concentrate on the magnetic fields
remaining after the end of the electroweak phase transition. We assume the
conversion from $W^0$ and $Y$ into photons to be efficient and not to
affect the resulting magnetic field distribution significantly so that we
may simply consider the coupling of $\phi$ to the photon field.

The axial coupling is characterised by the scalar function $f(\phi)$. We will see
later how this function affects the evolution of the vector potential. Notice that
if $f(\phi)$ was a constant, the EM part of the action would still be conformally
invariant and therefore, no EM fluctuations could be amplified during inflation.
Note also that the term $\mathcal{L}_I$ either breaks parity explicitly if $\phi$ is a
normal scalar field or, if $\phi$ is a pseudo-scalar, parity is broken spontaneously
by the presence of a background field $\phi\not\equiv 0$. For the discussion in this
work, this distinction is not relevant. However, in certain models, it might be
relevant for the amount of parity violation generated during reheating.

Varying the action with respect to $\phi$ leads to a sourced equation of motion
for the scalar field
\be
  \nabla^\alpha\partial_\alpha\phi - V'(\phi)
  \ =\ \frac{1}{4} f'(\phi)\, F_{\alpha\beta} \Fdual^{\alpha\beta} \,.
  \label{eq:fieldeq_scalar}
\ee
The primes in $V'(\phi)$ and $f'(\phi)$ denote derivatives with respect to $\phi$.
The field equations for the EM  field follow from varying the action with respect
to $A_\mu$:
\be
  \nabla_\alpha F^{\mu\alpha}\ =\ f'(\phi)\, (\partial_\alpha\phi)\, \Fdual^{\mu\alpha}
  \,.  \label{eq:fieldeq_maxinhom}
\ee
Comparison with the usual inhomogeneous Maxwell equation leads us to interpret the
source term on the right hand side as an effective axial current\footnote{An axial
anomaly which also induces a source term of this form has been discussed in
Ref.~\cite{Giannotti:2008cv}.}. To obtain the above form of the inhomogeneous
Maxwell equation, we used the homogeneous Maxwell equation
\be
  \nabla_\alpha \Fdual^{\mu\alpha}\ =\ 0 \quad \Leftrightarrow \quad
  \nabla_{[\lambda} F_{\mu\nu]}\ =\ 0
\ee
which is equivalent to the Bianchi identity, $dF=0$.

%%%%%%%%%%%%%%%%%%%%%%%%%%%%%%%%%%%%%%%%%%%%%%%%%%%%%%%%%%%%%%%%%%%%%%%%%
\subsection{Background evolution}  \label{sec:bg}

To describe the universe during inflation, we work in a flat
Friedmann-Lema\^{\i}tre (FL) background metric characterised by the line element
\be
  \dd s^2\ =\ a^2(\tconf)(-\dtconf^2 +\delta_{ij}\dx^i\dx^j)
\ee
where $a$ is the scale factor and $\tconf$ is conformal time which is related
to cosmic time $\tau$ by $a\,d\tconf=d\tau$. Derivatives with respect to
conformal time are
denoted by a dot and the conformal Hubble parameter is $\H\equiv\dot a/a=aH$ where
$H\equiv(\da/d\tau)/a$ is the physical Hubble parameter.

We assume the scalar field to dominate the energy budget of the universe and to
drive inflation. We shall check later under what conditions the EM energy density
is negligible and this approach is justified. We decompose the scalar field into
a background value and a small perturbation:
$\phi(x^\mu)\equiv\varphi(\tconf)+\delta\phi(x^\mu)$.
At background level, \eqref{eq:fieldeq_scalar} reduces to the homogeneous evolution
equation for the scalar field
\be
  \ddot\varphi + 2\,\H\dot\varphi +a^2V'(\varphi)\ =\ 0 \,.
\ee
The evolution of the scale factor is determined by the background Friedmann
constraint equation
\be\label{e:Fried}
  3\,m_P^2\H^2\  =\ \frac{1}{2}\dot\varphi ^2 + a^2V(\varphi)\,.
\ee

During slow roll inflation, the potential of the inflaton field is dominating the
energy density of the universe and the first term on the right hand side of
\eqref{e:Fried} can be neglected. This is quantified by means of the slow roll
parameters (see e.g.~\cite{Durrer:2008aa})
\be\label{eq:def_eps}
  \epsilon\ \equiv\ \frac{m_P^2}{2}\left(\frac{V'}{V}\right)^2\ \ll\ 1
  \,, \qquad \left|\frac{m_P^2}{3}\frac{V''}{V}\right| \ \ll\  1 \,.
\ee
To first order in the slow roll parameters, the evolution equation of $\varphi$
and the Friedmann constraint can be reduced to~\cite{Durrer:2008aa}
\be
  \dot\varphi\ \simeq\ - \frac{a^2 V'}{3\H}\
    =\ \pm \sqrt{2\epsilon}\, \frac{a^2V}{3m_P\H}\
    \simeq\ \pm \sqrt{2\epsilon}\, m_P \H \,.  \label{eq:sr_phidot}
\ee
The sign of $\dot\varphi$ depends on the details of the inflation model, but here
we do not assume a specific form of the potential. Using also that
$\H\simeq |\tconf|^{-1}$ and that $\epsilon$ is roughly constant in the slow roll
regime, we can integrate this result to find
\be\label{e:phislow}
  \varphi\ \simeq\ \varphi_{\rm in} \mp \sqrt{2\bar\epsilon}\,
    m_P\ln( \tconf/\tconf_{\rm in} )
\ee
where $\varphi_{\rm in}\equiv\varphi(\tconf_{\rm in})$ is the initial value of
the inflaton, and $\bar\epsilon$ is the average value of $\epsilon$ in the slow
roll regime. However, in this approximation, any deviation of $\epsilon$ from a
constant value is integrated over time, which can lead to significant deviations
in the evolution of $\varphi$ towards the end of inflation.

%%%%%%%%%%%%%%%%%%%%%%%%%%%%%%%%%%%%%%%%%%%%%%%%%%%%%%%%%%%%%%%%%%%%%%%%%
\subsection{Linear perturbation equations}

The EM field does not contribute to the background expansion but comes into play
at the perturbative level. We study the generation of perturbations in the FL
background during inflation. We work in longitudinal gauge where the metric
perturbation is
\be
  \delta g_{\mu\nu}\dx^\mu\dx^\nu\ =\ a^2(\tconf)\left(-2\Psi\dtconf^2
- 2\Phi\delta_{ij}\dx^i\dx^j\right)
\ee
and the two scalar degrees of freedom, $\Psi$ and $\Phi$, coincide with the
gauge-invariant Bardeen potentials~\cite{Durrer:2008aa}.

We work in Coulomb gauge throughout, i.e.~$A^\mu=(0,A^i)$ with $\partial_iA^i=0$.
To lowest order\footnote{Since the EM energy density is quadratic in the fields,
one considers the vector potential and the electric and magnetic fields to be at
half order in linear perturbation theory.}, the inhomogeneous Maxwell equation 
(\ref{eq:fieldeq_maxinhom}) with the axial current becomes
\be
  \ddot A_i -\nabla^2 A_i\ =\ -f'(\varphi)\dot\varphi\, \epsilon_{ijk}\partial_j A_k \,.
  \label{eq:lin_maxinhom}
\ee
where the Euclidean Laplacian is defined as
$\nabla^2\equiv\delta^{ij}\partial_i\partial_j$ and the totally anti-symmetric
symbol $\epsilon_{ijk}$ satisfies $\epsilon_{123}=1$. Note that these equations
are like in Minkowski space, there is no coupling to the scale factor. For a
constant axial coupling, $f'(\varphi)=0$, the sourced Maxwell equation reduces
to the standard free wave equation and no fluctuations are amplified during
inflation.

The evolution of perturbations in the scalar field is also altered by the axial
coupling. At linear order, the scalar field equation~(\ref{eq:fieldeq_scalar})
acquires a source term
\bea
  \ddot{\delta\phi} + 2\H\dot{\delta\phi} -\nabla^2\delta\phi
    + a^2V''(\varphi)\delta\phi  -\dot{\varphi}(3\dot\Phi+\dot{\Psi})
    + 2a^2V'(\varphi)\Psi  \nonumber \\  \label{e:dephi}
  \hspace{2cm}=\ -a^{-2}f'(\varphi)\epsilon_{ijk}\dot{A}_i \partial_j A_k \,.
\eea
For any scenario where EM perturbations are generated, it is important to investigate
the effect of this source term on the generation of scalar perturbations and through
these on the primordial curvature perturbations. For instance, this has been studied
in the case of natural inflation \cite{Anber:2009ua}. We leave a general discussion of
such effects for future work~\cite{WeInPrep} and concentrate here on the generation of
magnetic fields.

%%%%%%%%%%%%%%%%%%%%%%%%%%%%%%%%%%%%%%%%%%%%%%%%%%%%%%%%%%%%%%%%%%%%%%%%%
\subsection{Physical properties of electromagnetism in the expanding universe}
\label{sec:em}

The four-vector potential is generally covariant and its evolution is independent of
the choice of coordinates. However, for an observer, the physical EM field manifests
itself in terms of electric and magnetic fields which are intrinsically frame dependent
quantities. Measured by an observer with four-velocity $u^\mu$, with $u^\alpha u_\alpha=-1$,
the electric and magnetic fields can be covariantly defined as~\cite{Barrow:2006ch}
\bea
  E_\mu\ =\ F_{\mu\alpha} u^\alpha
  \\
  B_\mu\ =\ \frac{1}{2}\eta_{\mu\alpha\beta\gamma} F^{\alpha\beta} u^\gamma
    \ =\ \Fdual_{\mu\alpha} u^\alpha \,.
\eea
These are both three-vector fields in the sense that they are orthogonal to the observer
velocity, $E_\alpha u^\alpha=0=B_\alpha u^\alpha$. In a perturbed FL metric an observer
has the four-velocity $u^\mu=a^{-1}(1,\V{0})+\order{1}$. As a consequence one finds
(in Coulomb gauge)
\bea
  \left(E_\mu\right)\ =\ \Big(0,\ -\frac{1}{a}\dot{A}_i \Big)
  \,,\qquad
  \left(B_\mu\right)\ =\ \Big(0,\ \frac{1}{a}\epsilon_{ijk}\partial_j A_k \Big)
\eea
up to first order. With respect to an orthonormal basis comoving with the observer (or
generally the Hubble flow), we define the Euclidean three-vector fields $\VE$ and $\VB$
through
\be
 \left( E_\mu\right)\ =\ a\Big(0,\ \VE \Big)\,,\qquad   
\left(B_\mu\right)\ =\ a\Big(0,\ \VB \Big) \,.
\ee
As discussed in Ref.~\cite{Subramanian:2009fu}, in a highly conducting plasma, magnetic
fields should scale as $a^{-2}$ with the expansion, while the electric field is damped
away. The three-vector fields $\VE$ and $\VB$ defined above show exactly this property.
We therefore rescale the fields by a factor of $a^2$ such that the effect of the expansion
is absorbed, i.e.~$\VBtilde\equiv a^2\VB$.  In terms of the vector potential, the rescaled
fields become
\bea
  \widetilde{E}_i\ =\ -\dot{A}_i
  \,,\qquad
  \widetilde{B}_i\ =\ \epsilon_{ijk}\, \partial_j A_k \,.
  \label{eq:BfromA}
\eea
Using these expressions and the field equations for $A_\mu$, one can derive Maxwell's
equations for the rescaled fields $\VEtilde$ and $\VBtilde$, which take the same form
as in a Minkowski space-time.

The physical properties of the EM fields can now be described in terms of the rescaled
fields. Here, we are mainly interested in the energy density and the helicity density
of the fields generated during inflation. The energy density of the magnetic field is
\bea
  \rho_B\ \equiv\ \frac{1}{2}B_\alpha B^\alpha\ =\ \frac{1}{2}\VB\!\cdot\!\VB
  \ =\ \frac{1}{2}a^{-4}\,\VBtilde\!\cdot\!\VBtilde\ \equiv\ a^{-4}\,\widetilde{\rho}_B
\eea
and analogously for the electric energy density. The helicity density is given as
\be
  \hel\ \equiv\ A_\alpha B^\alpha\ =\ a^{-3}\,\delta^{ij}A_i\widetilde{B}_j
    \ \equiv\ a^{-3}\,\widetilde{\hel} \,.
\ee
(Deliberately, we are not denoting the vector-potential in bold face because
$A_i$ are the spatial components of the four-co-vector $A_\mu$ in the covariant
representation with respect to the coordinate basis $(dx^\mu)$, while
e.g.~$\VB$ are the components of the four-co-vector $B$ with respect to
the orthonormal basis $(adx^\mu)$ as discussed,
for instance, in Ref.~\cite{Subramanian:2009fu}.)

%%%%%%%%%%%%%%%%%%%%%%%%%%%%%%%%%%%%%%%%%%%%%%%%%%%%%%%%%%%%%%%%%%%%%%%%%
\section{Electromagnetic quantum fluctuations} \label{sec:modeeq}
%%%%%%%%%%%%%%%%%%%%%%%%%%%%%%%%%%%%%%%%%%%%%%%%%%%%%%%%%%%%%%%%%%%%%%%%%

To investigate the generation of EM fields during inflation, we consider the evolution
of quantum fluctuations of the EM vector-potential. The coupling of the vector-potential
to the background evolution of the inflaton via the inhomogeneous Maxwell
equation~(\ref{eq:lin_maxinhom}) can lead to the amplification of EM quantum fluctuations.
The amplification depends on the axial coupling $f(\varphi)$ and therefore on the
evolution of $\varphi$. In \ref{app:quantisation}, we review the quantisation of the
vector-potential in an expanding background. The result is that Maxwell's equations lead
to an evolution equation for the Fourier modes of the quantised field which reduces to
the free wave equation in the absence of the axial coupling. We first discuss this
evolution equation and then summarise the physical observables of the EM field expressed
in terms of the solutions to the mode equations.

%%%%%%%%%%%%%%%%%%%%%%%%%%%%%%%%%%%%%%%%%%%%%%%%%%%%%%%%%%%%%%%%%%%%%%%%%
\subsection{Evolution of the Fourier modes of the vector potential}

We introduce the orthonormal spatial basis as
\be
  \left(\Veps^\Vk_{1}\,,\,\Veps^\Vk_2\,,\,\hat\Vk\right) \ \qquad \mbox{with }
\quad  |\Veps^\Vk_{i}|^2=1 \,, \hat\Vk=\Vk/k\,,
\ee
and
\be
  \Veps^\Vk_{\pm}\ \equiv\
    \frac{1}{\sqrt{2}}\left( \Veps^\Vk_1 \pm i\Veps^\Vk_2 \right) \,.
\ee
In radiation gauge, the vector potential then takes the form
\be
  \VA\ =\ \A_1\Veps_1+\A_2\Veps_2\ =\ \A_+\Veps_++\A_-\Veps_- \,.
\ee
After quantisation of the vector potential, we can study the evolution of the Fourier
modes, $\A_h(\tconf,k)$, with respect to the helicity basis for the polarisation
states, $h=\pm$. We find that the helicity modes satisfy the wave equation with a time
dependent mass term corresponding to the modified Maxwell equation~(\ref{eq:lin_maxinhom})
for the classical vector-potential,
\be
  \ddot\A_h +\left[k^2+hkf'(\varphi)\dot\varphi\right]\A_h\ =\ 0 \,.
  \label{eq:modeeq}
\ee
The fact that the sign of the respective helicity mode, $h=\pm$, appears explicitly
in the evolution equation of the Fourier modes, leads to a different evolution of
the two helicity states and therefore, to the generation of magnetic helicity.
Also note that the scalar field couples to $k\A_h$ and the coupling function itself,
$f'(\varphi)\dot\varphi$, only depends on time, as $\varphi$ is the background value
of the inflaton. The solutions to this mode equation for a given coupling $f(\varphi)$
fully determines the spectrum of the generated EM fields.

Let us compare the mode equation~(\ref{eq:modeeq}) to the non-helical case with the
coupling $f(\varphi)F^2$. One obtains a similar evolution equation~\cite{Subramanian:2009fu}
\be
  \ddot\A_h +2\frac{f'\dot\varphi}{f}\dot\A_h +k^2\A_h\ =\ 0 \,.
  \label{eq:modeeqnonh}
\ee
Redefining the mode functions as $\bar{\A}_h\equiv f\A_h$, one can rewrite this in
the form
\be
  \ddot{\bar{\A}}_h +\left[k^2 -\frac{\ddot{f}}{f}\right]\bar{\A}_h\ =\ 0 \,.
  \label{eq:modeeqnonh2}
\ee
We observe two significant differences to the helical case: firstly, the two helicity
states couple with the same sign and therefore, no helicity is generated. Secondly, the
scalar field couples to $\dot\A_h$ (or $\bar\A_h$) as opposed to $k\A_h$ in the helical
case. As we shall see, this leads to significant differences in the spectra obtained
for the two types of couplings. The reason is that the additional factor $k$ in the
helical case leads to a suppression of the coupling term at super-Hubble scales.

\begin{figure}[ht]
\begin{center}
\includegraphics[width=0.65\textwidth]{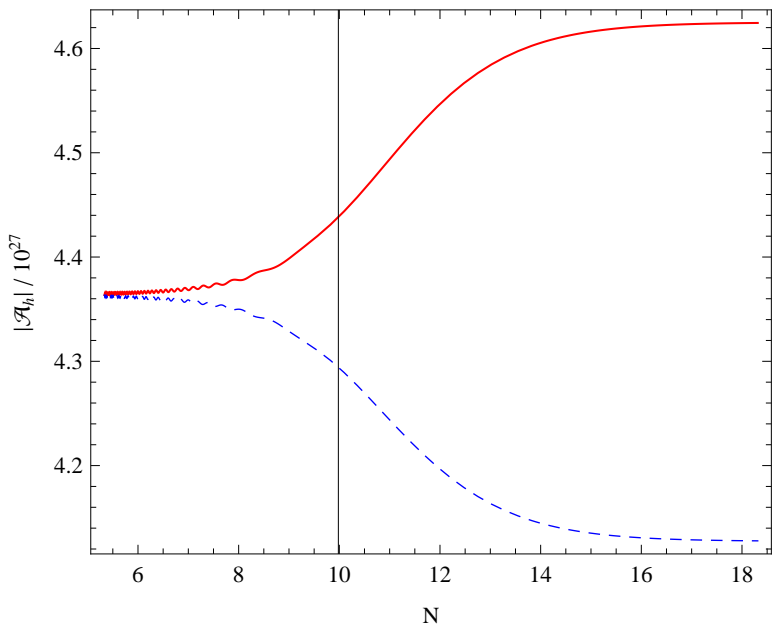}
\caption{The evolution of the two helicity modes ($+$ solid red, $-$ dashed blue) for
wavenumber $k=10/\Mpc$ is shown as a function of the number of e-foldings, $N$, during
inflation. Here we consider an axial coupling function $f\propto\varphi^p$, as discussed
in detail in \secref{sec:powcoup}. Both modes feel the axial coupling only around horizon
crossing, at $N_{\rm cross}\simeq 10$, while the evolution ceases and the modes saturate
quickly after crossing. Here inflation ends at $N\simeq 60$.}
\label{fig:singlemode}
\end{center}
\end{figure}

To visualise this, let us compare the importance of the different terms in the helical
mode equation~(\ref{eq:modeeq}). We define $\fN$ to be the logarithmic derivative of
the coupling function, $f$, with respect to the scale factor
\be
  \fN\ \equiv\ \frac{\dd f}{\dlna}\ =\ \frac{f'(\varphi)\dot\varphi}{\H}
\  \simeq \ \pm f'(\varphi)\sqrt{2\ep}\, m_P  \,.
\ee
where $N$ stands for the number of e-foldings and is defined as $N=\ln(a/a_{\rm in})$.
Note that $\fN$ is the dimensionless part of the coupling term appearing in
the mode equation (\ref{eq:modeeq}). First consider super-Hubble scales, $k
\ll \H$, where we can approximate $|\partial_\tconf|\sim\H$ so that
(\ref{eq:modeeq}) reduces to
\be
  \left[ 1+\frac{k^2}{\H^2} +h\fN\frac{k}{\H} \right] \A_h\ \simeq\ 0 \,.
\ee
The second term is small by definition and the coupling term can be important on
super-Hubble scales only if $f$ is a rapidly varying function of time,
i.e.~$\fN=\dot{f}/\H \gg 1$ and then only for as long as $\fN>\H/k$. On sub-Hubble
scales, for $k\gg\H$, we may approximate $\ddot\A_h \sim k^2\A_h$. As long as
$k\gg\fN\H$ the first and second terms in~\eqref{eq:modeeq} dominate. Hence the
coupling term is typically relevant only at Hubble crossing $k\sim \H$ during a
few Hubble times at best. In \figref{fig:singlemode}, we show the evolution of
the two helicity modes for a given wavenumber in case of a power law coupling,
$f\propto \varphi^p$. Clearly, the modes only feel the axial coupling around
horizon crossing. This fact turns out to be relevant for the resulting spectrum.
We will discuss this issue in more detail in \secref{sec:powcoup}.

%%%%%%%%%%%%%%%%%%%%%%%%%%%%%%%%%%%%%%%%%%%%%%%%%%%%%%%%%%%%%%%%%%%%%%%%%
\subsection{Vacuum solutions and initial conditions}

To study the generation of perturbations during inflation, we need to set the initial
conditions when the mass term $\pm\,k\H\fN$ in \eqref{eq:modeeq} can be neglected,
i.e.~well inside the Hubble horizon at early times. This condition is usually formulated
in terms of the variable $x\equiv -k\tconf$ which approaches infinity in this limit.
During slow roll inflation $\H\simeq-1/\tconf$, so that $x = -k\tconf\simeq k/\H\gg 1$
if the mode with wavenumber $k$ is well inside the Hubble horizon.

From the mode equation~(\ref{eq:modeeq}), we see that if initially
\be
  \fN(\varphi_{\rm in})\
  =\ f'(\varphi_{\rm in})\dot\varphi_{\rm in} /\H_{\rm in}\ \ll\ k/\H_{\rm in}
  \label{e:vacon}
\ee
the axial coupling term can be neglected with respect to the $k^2$ term and the mode
equation becomes a free wave equation. Its solutions are plane waves. We match to the
incoming vacuum solution described in~\ref{app:quantisation},
\be
  \A_h(\tconf,k)\ =\ \A^{\rm free}(\tconf,k)\ =\ (2k)^{-1/2}e^{-ik\tconf}
  \,, \qquad \mbox{for }\ \ -\!k\tconf\gg 1 \,.  \label{eq:Afree}
\ee
This is used in the following as initial condition for the solutions of the full
mode equation. Notice that the free plane wave solution only yields a valid initial
condition if \eqref{e:vacon} is satisfied.

%%%%%%%%%%%%%%%%%%%%%%%%%%%%%%%%%%%%%%%%%%%%%%%%%%%%%%%%%%%%%%%%%%%%%%%%%
\subsection{Power spectra and physical quantities}\label{sec:ps}

The statistical distribution of the EM fields as seen by an observer can now be
quantified in terms of a given solution for the helicity modes of the EM vector
potential. We define the magnetic power spectrum and relate it to the magnetic
energy and helicity density.

If the magnetic field generated by some process is statistically
homogeneous and isotropic, its spectrum is determined by two scalar functions
$P_S(k)$ and $P_A(k)$. Since the magnetic field is a divergence-free vector field
the two-point function of the Fourier components of the magnetic field can
be written as
\bea
  \langle \widetilde{B}_i(\tconf,\Vk) \widetilde{B}^*_j(\tconf,\Vq) \rangle
  \ =\ \frac{(2\pi)^3}{2} \delta(\Vk-\Vq)
  \Big\{(\delta_{ij}& -\hat{k}_i\hat{k}_j) P_S(\tconf,k) &
  \\ \nonumber
  & - i \epsilon_{ijn} \hat{k}_n P_A(\tconf,k) \Big\} &
\eea
where $P_S$ and $P_A$ are the symmetric and anti-symmetric parts of the power
spectrum, respectively. The symmetric part of the spectrum determines the energy
density while the   anti-symmetric part corresponds to the helicity density:
\be
  \langle \widetilde{B}_i(\tconf,\Vk) \widetilde{B}^*_i(\tconf,\Vq) \rangle
  \ =\  (2\pi)^3\delta(\Vk-\Vq)P_S(\tconf,k)
\ee
\bea
 \langle \widetilde{A}_i(\tconf,\Vk) \widetilde{B}^*_i(\tconf,\Vq) \rangle
  &\ =\ i k^{-2}\langle (\Vk\wedge \widetilde{\VB})_i(\tconf,\Vk)
    \widetilde{B}^*_i(\tconf,\Vq) \rangle &     \nonumber \\
  &\ =\ k^{-1}(2\pi)^3 \delta(\Vk-\Vq)P_A(\tconf,k) \,. &
\eea
With respect to the helicity basis, see~\ref{app:quantisation}, the spectra can
directly be written as
\be\label{e:PSA}
  P_{S/A}(\tconf,k)\ =\ k^2\left( |\A_+(\tconf,k)|^2 \pm |\A_-(\tconf,k)|^2 \right) \,,
\ee
where the upper sign corresponds to $P_S$ and the lower sign to $P_A$. Here we use
the non-trivial result widely applied in inflationary cosmology that at late times,
the vacuum expectation values of the fields generated during inflation can be
interpreted as stochastic power spectra. 

We define the magnetic energy density per logarithmic wave number via
\be\label{e:avg}
  \langle \widetilde{\rho}_B(\tconf) \rangle\ =\ \int_0^\infty\frac{\dk}{k}\,
  \frac{\dd \widetilde{\rho}_B}{\dlnk}(\tconf,k) \,,
\ee
so that
\be
  \frac{\dd \widetilde{\rho}_B}{\dlnk}(\tconf,k)\ =\ \frac{k^3}{(2\pi)^2}
  P_S(\tconf,k)\,.
\ee
Similarly, we define the magnetic helicity per logarithmic wave number as
$\widetilde{\hel}=A_\alpha \widetilde{B}^\alpha$,
\be
 \frac{\dd\widetilde{\hel}}{\dlnk}(\tconf,k)\ =\frac{k^2}{2\pi^2}P_A(\tconf,k) \,.
\ee

Finally, the electric field is given by the time derivative of the vector-potential
and thus its contribution to the energy density,
$\widetilde{\rho}_E=\widetilde{E}_\alpha\widetilde{E}^\alpha /2$, is computed to be
\be
  \frac{\dd \widetilde{\rho}_E}{\dlnk}(\tconf,k)
  \ =\ \frac{k^3}{(2\pi)^2} \left( |\dot\A_+(\tconf,k)|^2 + |\dot\A_-(\tconf,k)|^2 \right)\,.
\ee
Notice that any electric fields produced during inflation will be damped very rapidly
after inflation due to the huge conductivity of the primordial plasma.

%%%%%%%%%%%%%%%%%%%%%%%%%%%%%%%%%%%%%%%%%%%%%%%%%%%%%%%%%%%%%%%%%%%%%%%%%
\section{Analytic solutions during slow roll inflation} \label{sec:slowroll}
%%%%%%%%%%%%%%%%%%%%%%%%%%%%%%%%%%%%%%%%%%%%%%%%%%%%%%%%%%%%%%%%%%%%%%%%%

In this section, we first derive a condition on the coupling function $f(\varphi)$ such
that the theory can be treated perturbatively. Even though we cannot prove that the theory
does not make sense otherwise, our treatment is perturbative and so we can really trust it
only if interactions are small. In our subsequent analysis in
Section~\ref{sec:interpretation} we shall, however, analyse the results also if this
condition is not satisfied. We then investigate two different forms of the coupling
function and obtain upper bounds on its parameters without specifying a model of inflation.
Finally, we solve the mode equation in slow roll for a constant $\fN$ analytically, derive
the resulting magnetic field power spectrum at the end of inflation and discuss the issue
of backreaction on the background evolution of the inflaton.

%%%%%%%%%%%%%%%%%%%%%%%%%%%%%%%%%%%%%%%%%%%
\subsection{Condition on the coupling function}  \label{sec:coupf}
%%%%%%%%%%%%%%%%%%%%%%%%%%%%%%%%%%%%%%%%%%%

In order to derive a condition on the coupling function such that the theory can be treated
perturbatively, the interaction between the scalar field and the EM field should be small at
all times. For this we demand that the ratio of the actions for the interaction term to
the free EM term should be less than unity, i.e.
\be\label{e:comp}
  \left|\frac{S_I[\phi,A_\mu]}{S_{\rm em}[A_\mu]}\right|\ <\ 1.
\ee
In \ref{app:perturbativity} we show that in a FL background the free EM action can be
written as
\be
  S_{\rm em}[A_\mu]\ =\ V \int d\tconf\, \langle \widetilde{\rho}_E(\tconf) -
  \widetilde{\rho}_B(\tconf)\rangle_V
\ee
where $\langle\ldots\rangle_V$ denotes the average over volume $V$ in coordinate space 
and is assumed to be equivalent to the expectation value defined in~\eqref{e:avg}.
Similarly, the interaction part of the action can be expressed in terms of the magnetic helicity
as (see~\ref{app:perturbativity})
\be
  S_I[\phi, A_\mu]\ =\ \frac{1}{2}V\int d\tconf\, \dot f\, \langle 
\widetilde \hel(\tconf)\rangle_V \ .
\ee
To compare the two parts of the action, one has to evaluate the time integrals for a 
given scenario.
For the case of a maximally helical magnetic field, e.g. $|\A_{+}|=|\A|$ with $|\A_{-}|=0$
and assuming a power law form for the spectrum given by
\be
  P_S\ =\ P_A\ =\ k^2 |\A|^2\ =\ B_{0}^2\,k^n \,,
\ee
we find that
\bea
  \langle \widetilde{\rho}_B(\tconf)\rangle_V 
  &\ \simeq\ \frac{B_{0}^2}{(2\pi)^2}\, \frac{k_{\rm max}^{n+3}(\tconf)}{(n+3)} \\
  \langle \widetilde \hel(\tconf) \rangle_V 
  &\ \simeq\ \frac{2B_{0}^2}{(2\pi)^2}\, \frac{k_{\rm max}^{n+2}(\tconf)}{(n+2)}
\eea
where $k_{\rm max}(\tconf)$ is the cut-off scale which is the smallest 
scale crossing the Hubble scale at time $\tconf$ when the volume
average is computed, hence $k_{\rm max}\simeq \H(\tconf)$. To estimate the energy
density for the electric field, we can approximate $|\dot \A| \simeq \H |\A|$ at 
large scales which then leads to
\be
  \langle \widetilde{\rho}_E(\tconf)\rangle_V
  \ \simeq\ \frac{B_{0}^2}{(2\pi)^2}\, \frac{k_{\rm max}^{n+1}(\tconf)}{(n+1)} \H^2
\ee
After evaluating the volume averages of the two energy densities and the helicity, we can
now compare the two actions. Ignoring numerical factors, the comparison condition~(\ref{e:comp})
can be written as
\be
  \left|\frac{S_I[\phi,A_\mu]}{S_{\rm em}[A_\mu]}\right|\ \simeq\
 \left|\frac{\int d\tconf\, \dot f\,\H^{n+2}}{\int d\tconf\,\H^{n+3}}\right|\ <\ 1.
\ee
Now, for the above inequality to be satisfied, it is sufficient to require $|\dot f|<\H$ or
equivalently $|\fN|<1$.
We believe that this condition is quite generic and does not depend on the functional
form of the coupling function. The above analysis indicates the fact that the condition
for the perturbative treatment of the theory does not depend on the coupling function,
but on its derivative.
This is not surprising as a constant $f$ only yields a surface term which does not affect the
dynamics.

In flat spacetime the above condition has to be replaced by $|\dot f|<k_{\rm max}$, where
$k_{\rm max}$ denotes the UV cutoff of the modes under consideration.

%%%%%%%%%%%%%%%%%%%%%%%%%%%%%%%%%%%%%%%%%%%
\subsection{Power law and exponential coupling}  \label{sec:powcoup}
%%%%%%%%%%%%%%%%%%%%%%%%%%%%%%%%%%%%%%%%%%%

We first study the special case of a power law coupling,
\be\label{eq:fpow}
  f(\varphi)\ =\ f_0 \left(\frac{\varphi}{m_P}\right)^p \,.
\ee
Here $f_0$ and $p$ are constants. The sign of $f_0$ determines which helicity is amplified
by the coupling and, thus, we can take $f_0$ to be positive without loss of generality.
Let us first investigate bounds on the parameter space of $f_0$ and $p$ under the condition
$|\fN| < 1$.

During slow roll inflation, we use \eqref{eq:sr_phidot} to express the coupling term in the
mode equation~(\ref{eq:modeeq}), $\fN=f'(\varphi)\dot\varphi/\H$, in terms of the slow roll
parameter $\epsilon$
\be
  \fN\ \simeq\ \pm\, f_0\, p\, \sqrt{2\epsilon} \left(\frac{\varphi}{m_P}\right)^{p-1}
\ee
where, in principle, $\epsilon$ is a slowly varying function of time. If we want to
satisfy  the condition of perturbativity,
$|\fN| < 1$, the above equation leads to an upper bound on the value of $f_0$ as
\be
  f_0\ <\ f_{0}^{\rm max}\ =\ \frac{1}{|p|} \min\limits_\varphi
    \left\{\frac{1}{\sqrt{2\epsilon}}\left(\frac{m_P}{\varphi}\right)^{p-1}\right\}.
  \label{eq:f0max_pow}
\ee
For a given inflationary scenario, one can invert the definition of $\epsilon$ to express the
above bound only as a function of $\varphi$. Though it turns out to be easy for large field
inflationary models, it is not so straightforward in the case of typical models of small field
inflation. Once the upper bound on $f_0$ is known, one can calculate the maximal possible
value of the coupling term i.e.~$\fNmax(p) \equiv \fN(f_{0}^{\rm max}(p),p)$.

As a second case, we consider the example of an exponential chiral coupling of the 
following form
\be
  f(\varphi)\ =\ f_0\, e^{\alpha\, \varphi/{m_P}}
\ee
where $f_0>0$ and $\alpha$ are constants.
Analogously to the power law case above, we calculate the coupling term $\fN$ in terms of
$\epsilon$ which is given by
\be
  \fN\ \simeq\ \pm\,f_{0}\,\alpha \sqrt{2\epsilon}\; e^{\alpha\, \varphi/{m_P}}
\ee
We again apply the constraint $|\fN|<1$ to find the bound on $f_0$:
\be
  f_0\ <\ f_{0}^{\rm max}\ =\ \frac{1}{|\alpha|} \min\limits_\varphi
  \left\{\frac{1}{\sqrt{2\epsilon}}\,e^{-\alpha\, \varphi/{m_P}}\right\}.
  \label{eq:f0max_exp}
\ee
As before, once $f_{0}^{\rm max}$ is known, the maximum value of the coupling term
$\fNmax$ can be calculated.

In both models, the power law and the exponential chiral coupling, the constraint
$|\fN|<1$ leads to an upper bound on the overall amplitude of the coupling. The time
dependence of $\fN$, on the other hand, depends on the details of the inflation model.
Though, generally we remark that for a given mode $k$ the coupling is only
active for a few e-foldings
during which the mode crosses the Hubble scale, and moreover, if $\fN$ had a
significant time variation its overall amplitude would be suppressed very strongly
by the bound on $f_0<f_0^{\rm max}$. Therefore, we conclude that $\fN$ can safely
be considered roughly constant (at least during horizon crossing).

%%%%%%%%%%%%%%%%%%%%%%%%%%%%%%%%%%%%%%%%%%%%%%%%%%%%%%%%%%%%%%%%%%%%%%%%%
\subsection{Analytic solution for a constant coupling term}

We now  solve the mode equation~(\ref{eq:modeeq}) for the case where $\fN$ is
approximately constant. 
The solution presented here will be valid for any constant value of $\fN$. 
In the slow roll regime, the full coupling term is,
$f'(\varphi)\dot\varphi=\fN\H\simeq -\fN/\tconf$, i.e.~inversely proportional
to conformal time and, consequently, the mode equation can be solved analytically.
It is convenient to use $x\equiv -k\tconf$ as the time evolution variable. For
each scale, $k$, the initial condition in the asymptotic past is set well inside
the horizon, i.e.~for $x\to\infty$, while inflation is considered to end when
$x\to 0$. With a prime denoting the derivative with respect to $x$, the mode
equation reads
\be\label{e:modeClb}
  \A_h''(x) +\left(1+h\fN x^{-1}\right)\A_h(x)\ =\ 0 \,.
\ee
The free solution is a good approximation at early times, $x\gg |\fN|$.
The general solution to the mode equation~(\ref{e:modeClb}) is~\cite{Abramowitz:1972}
\be
  \A_h(x)\ =\ C_1 G_0(y,x) + C_2 F_0(y,x)
\ee
where $y\equiv -h\fN/2$ and $G_0$ and $F_0$ are the irregular and regular Coulomb
wave functions of order zero, respectively. As initial condition we require the
solution to approach the free solution, \eqref{eq:Afree}, for $x\to\infty$. It
turns out that the combination $G_0\pm iF_0$ has the desired limit, as given in
Ref.~\cite{Abramowitz:1972}:
\be
  G_0(y,x)\pm iF_0(y,x)\ \mathop{\longrightarrow}_{x\to\infty}\
  \exp{\pm i\left[x-y\ln(2x) +\sigma_0(y)\right]}
\ee
with $\sigma_0(y)\equiv\arg\Gamma(1+iy)=\gamma_E y+\order{y^3}$ and
$\gamma_E\simeq 0.58$ being the Euler's constant. For large $x$, the second term in
the exponential, $y\ln(2x)$, can be neglected
with respect to $x$. Comparison with the free solution shows that the plus sign
corresponds to the incoming vacuum solution and we have to choose the initial amplitude
\be
  \frac{1}{2}(C_1-iC_2)\ =\ (2k)^{-1/2}
\quad \mbox{ and } \qquad  \frac{1}{2}(C_1+iC_2)\ =\ 0 \,.
\ee
The factor with $\sigma_0$ only acts as an overall phase and has no physical significance.
The normalised full solutions are
\be
  \A_h(x)\ =\ (2k)^{-1/2}\left[ G_0(-h\fN/2,x) + iF_0(-h\fN/2,x) \right]
\ee
for $h=\pm 1$. This solution of the mode equation has also been found in
Ref.~\cite{Anber:2006xt} for $N$-flation. Below we shall argue that it is very general.

To understand the effect of the axial coupling on the growth of EM quantum fluctuations
and to compute the magnetic power spectra at the end of inflation, we analyse the late
time limit of this solution, i.e.~when $x\to 0$. Using the approximate expressions
derived in \ref{app:coulomb}, we find the asymptotic limit at late times as
\be
  \A_h(x)\ \mathop{\longrightarrow}_{x\to 0}\ (2k)^{-1/2}
  \left[ \frac{\exp(-h\pi\fN/2) \sinh(\pi\fN/2)}{\pi\fN/2} \right]^{1/2} \,.
\ee
The late time behaviour of both helicity modes is independent of $x$, and therefore of
$\tconf$, and the scale-dependence is not changed with respect to the free solution.
The modes are coherently amplified while crossing the horizon, before they saturate
outside the horizon. Notice also that we did not use the asymptotic limit given in
$14.6.9-10$ of Ref.~\cite{Abramowitz:1972} because it is not correct. (In this context,
see discussion in \ref{app:coulomb}.)

\begin{figure}[t]
\begin{center}
\includegraphics[width=0.65\textwidth]{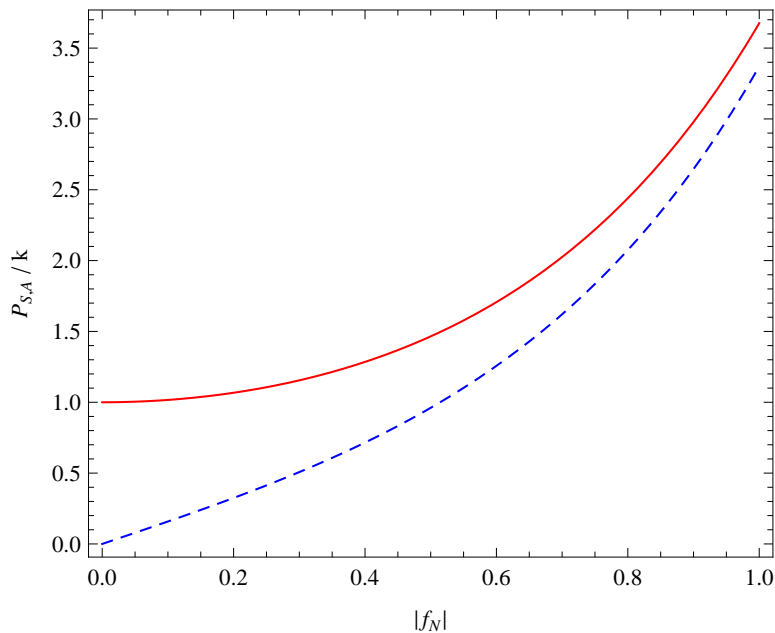}
\caption{The symmetric (solid, red) and antisymmetric (dashed, blue) power spectra 
of the magnetic fields, $P_{S,A}/k$, are plotted as a function of the effective 
coupling constant $|\fN|$. For vanishing coupling, the vacuum solution, $P_S/k=1$ 
and $P_A/k=0$ is recovered, while the larger the coupling,
the smaller the difference between $P_S$ and $P_A$. As discussed in the main text,
for the theory to be perturbative we must require $|\fN|<1$
and therefore, the amplification of the magnetic fields is small.}
\label{fig:spectral_amp}
\end{center}
\end{figure}

Given the late time behaviour of the helicity mode functions, it is easy to compute the
magnetic field power spectra produced at the end of inflation. The symmetric power
spectrum is
\be\label{e:PSres}
  P_S(k)\ = \ k\, \frac{\sinh(\pi\fN)}{\pi\fN}
  \label{eq:PSanalytic}
\ee
while the antisymmetric one is
\be\label{e:PAres}
  P_A(k)\ =\ k\, \frac{\cosh(\pi\fN)-1}{\pi\fN}
\ee
Notice that both spectra are proportional to $k$ and only their amplitude changes
with the coupling strength $\fN$. In \figref{fig:spectral_amp}, we illustrate the
$k$-independent amplification factors of $P_S$ and $P_A$ as a function of $\fN$.
The larger $|\fN|$ the smaller the difference between $P_S$ and $P_A$, i.e.~the
more helical the magnetic fields become. As one infers from Eqs.~(\ref{e:PSres})
and~(\ref{e:PAres}), both amplification factors tend to
$(2\pi\fN)^{-1}\exp(\pi\fN)$ for large values of $\fN$.

The magnetic energy density per logarithmic wave number at the end of inflation can directly
be computed to be
\bea \label{e:rhoM}
  \frac{\dd \rho_B}{\dlnk}(\tconf_{\rm end},k)
       \ = \ \frac{k^4}{a_{\rm end}^4}\,\frac{\sinh(\pi\fN)}{4\pi^3\fN}
     \ \equiv \ \frac{k^4}{a_{\rm end}^4}\,\CS^2(\fN)\,.
\eea
Here, we define $\CS^2(\fN)$ to be the amplitude of magnetic energy density spectrum. Since
its spectrum is blue, the magnetic energy density is dominated by the cut-off scale, which
is set by the last scale that exits the horizon before the end of inflation,
$k_c=\H_{\rm end}=(aH)_{\rm end}$. With this we obtain
\be
  \rho_B(\tconf_{\rm end})\ \simeq\ \frac{1}{4}H_{\rm end}^4\,\CS^2(\fN) \,.
\ee
By means of the Friedmann equation, we have
\be
  \Omega_B(\tconf_{\rm end})\ \equiv\ \frac{\rho_B}{\rho_{\rm tot}}
  \ \simeq\ \frac{\CS^2(\fN)}{12} \left(\frac{H_{\rm end}}{m_P}\right)^2\,.
\ee
The generic condition for backreaction to be negligible then is
\be\label{e:back}
  \CS(\fN)\ \lesssim\ \frac{m_P}{H_{\rm end}} \,.
\ee
If we assume that the reheating phase is short so that
$3m_P^2H_{\rm end}^2 = \rho_{\rm tot} \simeq T_*^4$, where $T_*$ is the reheating
temperature, we can write the above bound from backreaction as
\be\label{e:backT}
  \CS(\fN)\ \lesssim\ \frac{m^2_P}{T_*^2} \,.
\ee
For couplings that are perturbative in the naive sense discussed above, $|\fN|<1$, 
$\CS(\fN)$ is of order unity and, therefore, no backreaction on the background
evolution is expected if inflation ends well below the Planck scale.
However, the results above hold for any constant value of $\fN$, even if it was
larger than unity. Since we cannot prove that perturbativity is strictly required,
we shall therefore not restrict ourselves to this case in the following.
For instance, if the EM field was coupled to a large number, ${\mathcal N}$, of
pseudo-scalar fields then $\fN\to\sqrt{\mathcal N}\fN$, a situation where the effective
coupling term can be large without spoiling the perturbativity of the interactions
with the individual scalar fields, see Ref.~\cite{Anber:2006xt}.

In the more general case where $\fN$ is time dependent, in principle we have to evaluate
$\fN$ at horizon crossing. This would lead to a slight modification of the spectrum but
would not spoil the discussion of backreaction above. In the following we neglect this
effect, as it is quite irrelevant for the few orders of magnitude in $k$ which we are
interested in, see Fig.~\ref{fig:ps_dsr_pow}.

At the end of inflation and after reheating, we expect the universe to be filled with
relativistic standard model particles, a relativistic highly conducting plasma. In this
medium, the MHD approximation is valid and electric fields are rapidly damped away. We
therefore do not discuss the electric field spectrum which will not survive reheating.

The helicity density per logarithmic wave number at the end of inflation is simply given
by
\be\label{e:helM}
  \frac{\dd{\hel}}{\dlnk}(\tconf_{\rm end},k)
  \  = \ \frac{k^3}{a_{\rm end}^3}\, \frac{\cosh(\pi\fN)-1}{2\pi^3\fN} \,.
\ee
Note that $\widetilde\hel\propto\widetilde\rho_B/k_c$, which therefore has to be constant
when helicity is conserved.

We believe that this result is more general than the particular cases studied here: 
whenever the function $\fN$ is rapidly varying so that $\fN(\tconf)=$ constant 
during slow roll is no longer a good approximation, the fact that we require 
$|\fN|<1$ during the entire period of inflation means that $\fN$ must be oscillating. 
Though, because the coupling is active only for the small number of e-foldings
during which a mode crosses the Hubble scale, resonant amplification of modes at
horizon-crossing is not likely to be efficient. We have checked this statement 
numerically using different forms of oscillating coupling terms.
We, therefore, conclude that as long as the slow roll approximation is valid and the theory can be treated perturbatively, the amplification of helical magnetic fields is always mode independent and consequently leads to a $n=1$ spectrum for both the magnetic field and the helicity.

Note that this result differs significantly from the non-helical case. There, the source
term in the mode equation is of the form $\ddot f/f$ which is typically $\propto 1/\tconf^2$.
The solutions are then Bessel functions and the Bessel function index, which determines the
spectral index at late times, is related to the (nearly arbitrary) pre-factor. The difference
comes from the fact that a term of order $1/\tconf^2$ is relevant during all the time when
the mode is super-Hubble, $-k\tconf<1$, while a term of order $(k/\tconf)\A_h$ is relevant
only around horizon crossing. On super-Hubble scales, it is dominated by the
$\ddot \A_h \sim \A_h/\tconf^2$ term while on sub-Hubble scales, the $k^2 \A_h$ term becomes
dominant.

%%%%%%%%%%%%%%%%%%%%%%%%%%%%%%%%%%%%%%%%%%%%%%%%%%%%%%%%%%%%%%%%%%%%%%%%%%
\section{Deviations from slow roll}  \label{sec:numerical}
%%%%%%%%%%%%%%%%%%%%%%%%%%%%%%%%%%%%%%%%%%%%%%%%%%%%%%%%%%%%%%%%%%%%%%%%%%

In the previous section, we discuss two different functional forms of the axial
coupling of the inflaton to the EM field, namely a power law and an exponential.
Within the slow roll approximation we find that $\fN$ can safely be considered constant.
This always leads to a magnetic field power spectrum proportional to $k$. In this 
section, we first confirm our analytical findings numerically. Second, we 
explore the possibility of obtaining a different magnetic
field spectrum by introducing a short deviation from slow roll, motivated by the
fact that such deviations can provide a considerably better fit to the angular
power spectrum of the CMB anisotropies than the predictions from typical single
field inflation models, see for instance
Refs.~\cite{Covi:2006ci, Hamann:2007pa, Mortonson:2009qv, Jain:2008dw, Jain:2009pm, Hazra:2010ve}.

To compare the analytical result, \eqref{eq:PSanalytic}, to a full numerical solution,
we solve the background evolution of $\varphi$ with a quadratic potential,
$V_0(\varphi)\equiv \frac{1}{2}m^2\varphi^2$, and integrate the evolution of the
modes $\A_h(\tconf,k)$ to compute the magnetic power spectrum $P_S(k)$ at the end of
inflation.

A short deviation from slow roll can, for example, be achieved by introducing a step in the
quadratic inflaton potential as
follows~\cite{Covi:2006ci, Hamann:2007pa, Mortonson:2009qv, Adams:2001vc}
\be
  V_\beta(\varphi)\ =\ \frac{1}{2}m^2\varphi^2 \left[1+\beta\,{\rm tanh}
  \left(\frac{\varphi-\varphi_{0}}{\Delta\varphi}\right)\right] \,.
  \label{e:pot-step}
\ee
Here $\beta$, $\varphi_{0}$ and $\Delta\varphi$ characterise the height, the location and
the width of the step, respectively. Such a deviation from slow roll, in general, leads
to a burst of oscillations in the primordial power spectrum of curvature perturbations.
In what follows, we turn our attention to the possible effects of such a deviation from
slow roll on the magnetic field power spectrum.

\begin{figure}[t]
\begin{center}
\includegraphics[width=0.65\textwidth]{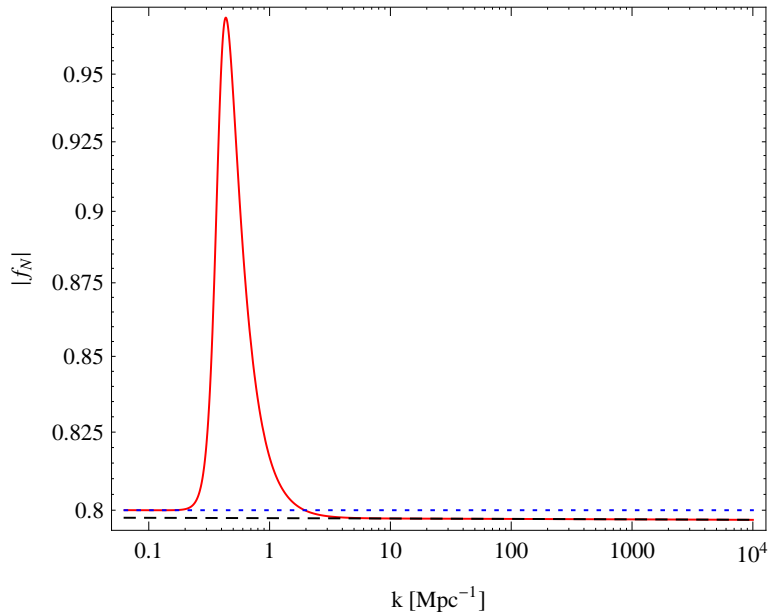}
\caption{The coupling term $|\fN|$ at Hubble crossing of the wave number $k$ is plotted
for the power law coupling with $p=2$, $f_0=1/5$. The horizontal dotted blue line
represents the constant value of $\fN=4/5$ in the slow roll approximation. The dashed
black line indicates the behaviour of the coupling term for the potential in
\eqref{e:pot-step} with $\beta=0$, while the solid red line is the same for
$\beta \neq 0$. For the later case, we have used the best fit values of the parameters
of the potential as in Ref.~\cite{Hazra:2010ve}. The bump in the coupling term for
$\beta \neq 0$ arises due to the short period of deviation from slow roll.}
\label{fig:dsr_pow}
\end{center}
\end{figure}

We perform the comparison with a power law coupling function, $f(\varphi)=
f_0(\varphi/m_P)^p$. In this case we find an exact expression for the slow roll 
parameter: $\sqrt{2\epsilon}=2m_P/\varphi$. Using this in \eqref{eq:f0max_pow} the 
upper bound on the parameter $f_0$ becomes
\be
  f_0^{\rm max}\ =\ \frac{1}{2|p|}
  \left\{ \begin{array}{cc} (\varphi_{\rm max}/m_P)^{2-p} &\quad {\rm for}\ p>2
  \vspace{2mm} \\
    (\varphi_{\rm min}/m_P)^{2-p} &\quad {\rm for}\ p<2 \end{array} \right.
\ee
and $\varphi_{\rm min}\leq \varphi \leq \varphi_{\rm max}$. With $f_0$ fixed to be 
$f_0^{\rm max}(p)$ the coupling term $\fNmax\equiv \fN(f_0^{\rm max},p)$ reads
\be
  \fNmax\ =\ \left\{ \begin{array}{cc}
  (\varphi/\varphi_{\rm max})^{p-2} &\quad {\rm for}\ p>2 \vspace{2mm} \\
  (\varphi/\varphi_{\rm min})^{p-2} &\quad {\rm for}\ p<2 \end{array} \right.
\ee
Thus, the most simple choice of parameters is: $p=2$ such that $f_0^{\rm max}=1/4$ 
and $\fNmax$ is exactly constant in slow roll. Furthermore, we choose 
$f_0=(4/5)f_0^{\rm max}=1/5$ to make sure that also in the case with the deviation 
from slow roll the condition $|\fN|<1$ is respected.

\begin{figure}[t]
\begin{center}
\includegraphics[width=0.65\textwidth]{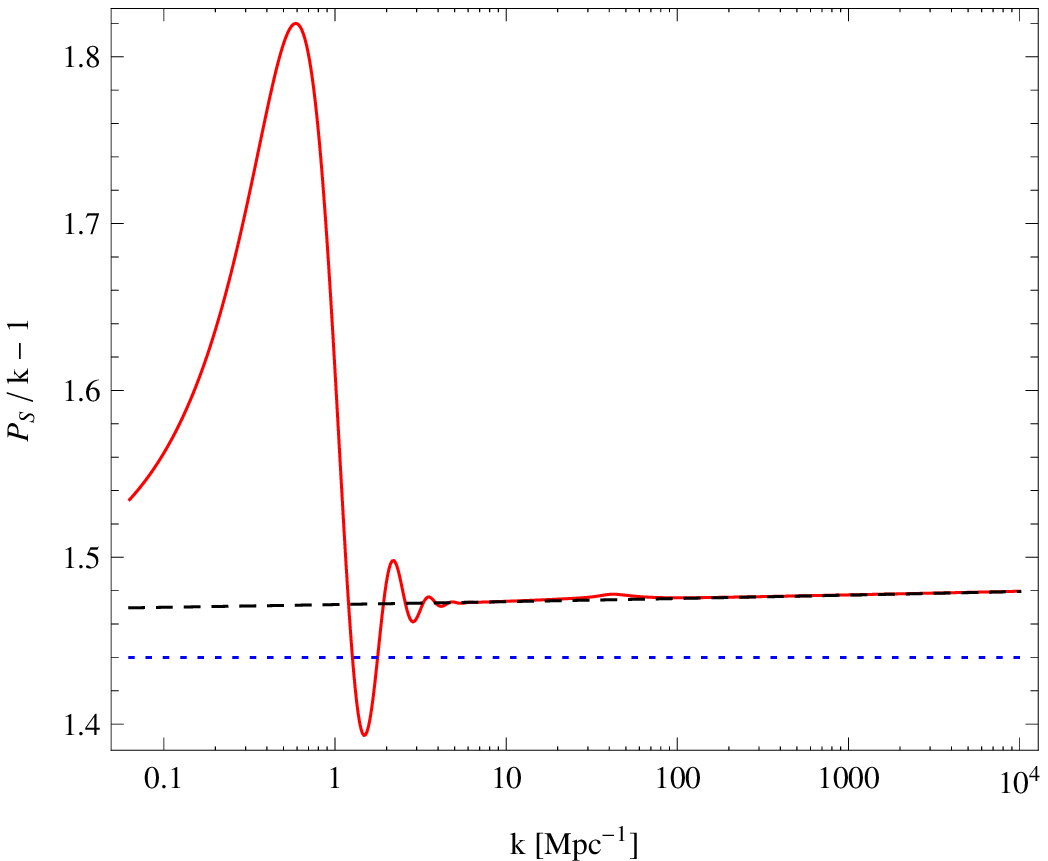}
\caption{The relative deviation of the magnetic field power spectrum from an exact
$k$-spectrum is plotted as a function of the wave number $k$ for the power law coupling
with $p=2$, $f_0=1/5$. The dashed black line indicates the numerical solution for
slow roll inflation ($\beta=0$) while the solid red line is the spectrum resulting from
a deviation from slow roll. The horizontal dotted blue line is the magnetic field
spectrum from the slow roll approximation, $\fN=4/5$. The deviation from this 
approximated spectrum is always small. Only the scales which exit the Hubble radius
around the time when a deviation from slow roll {\it or} equivalently a bump in the
coupling term occurs are affected more strongly.}
\label{fig:ps_dsr_pow}
\end{center}
\end{figure}

As we discuss earlier, the coupling term $\fN$ is typically relevant only around Hubble
crossing for a few Hubble times. The coupling term is, in general, a function of time but
this time dependence can be translated into a scale dependence by identifying a time with
the corresponding Hubble crossing scale, $k=\H(t)$. In \figref{fig:dsr_pow} we plot the
coupling term $\fN$ at Hubble crossing of the mode $k$ for the power law coupling with
$p=2$ and $f_0=1/5$. Comparing the slow roll value, $\fN=4/5$, with the
numerical results for the slowly rolling model and the case departing from slow roll, 
it is evident from the figure that a deviation from slow roll
leads to a bump in the coupling term as compared to the slow roll case and
therefore, one can expect an effect in the magnetic field power spectrum on the scales
which exit the Hubble radius around the time when the bump in the coupling term occurs.

In \figref{fig:ps_dsr_pow}, we plot the relative deviation of the magnetic field power
spectrum over an exact $k$-spectrum as a function of $k$ for the power law coupling. The
exact numerical solution for slow roll deviates slightly from the $P_S\propto k$ spectrum
due to the slight scale dependence of $\CS^2(\fN)$. Modulations in the spectrum arise
as a result of a deviation from slow roll. We find that for the best fit values of the
parameters of the potential~(\ref{e:pot-step}), the spectrum of the magnetic field is not
strongly modified. Indeed, we conclude that even the deviation from slow roll, within the
limits required by CMB data, does not significantly modify the magnetic field spectrum.
We find that the exponential coupling leads to a similar behaviour for the coupling term
and the magnetic field spectrum.

%%%%%%%%%%%%%%%%%%%%%%%%%%%%%%%%%%%%%%%%%%%%%%%%%%%%%%%%%%%%%%%%%%%%%%%%%
\section{The magnetic field at the end of inflation and its further evolution}
\label{sec:interpretation}
%%%%%%%%%%%%%%%%%%%%%%%%%%%%%%%%%%%%%%%%%%%%%%%%%%%%%%%%%%%%%%%%%%%%%%%%%

Our main results are Eqs.~(\ref{e:rhoM}) and~(\ref{e:helM}) which determine the magnetic
energy density and helicity density at the end of inflation. Note, however, that we do not
renormalise the energy density. Hence even for $\fN=0$, we obtain the non-vanishing result
\bdm
  \frac{\dd \widetilde\rho_B}{\dlnk}(\tconf_{\rm end},k) \ = \ \frac{k^4}{(2\pi)^2}
\edm
which comes purely from (not amplified) vacuum fluctuations and may be considered
unphysical. However, from \eqref{e:PSA} it is clear that $P_S(k)\geq |P_A(k)|$ by 
definition. Hence the physical result cannot be obtained by a simple subtraction 
of the vacuum contribution as then $P_S \propto \fN^2$ would become smaller than 
$P_A\propto \fN$ for small values of $\fN$,
see  Eqs.~(\ref{e:PSres}) and~(\ref{e:PAres}). On the other hand, for $\fN\gsim 1$, the
vacuum contribution becomes subdominant and it is no longer important to subtract it.
We shall therefore not perform any renormalisation of the magnetic energy density but just
keep in mind that our result becomes dominated by vacuum fluctuations in the limit $\fN\ra 0$.

After inflation, the thermal cosmic plasma contains many relativistic charged particles and
can be treated as an MHD plasma. During the process of reheating, the Reynolds number becomes
very high and MHD turbulence develops. In the MHD limit the electric field is damped away and
the magnetic field evolves by two different processes: it is damped on small scales and it
undergoes an inverse cascade due to helicity conservation~\cite{Brandenburg:2004jv}.
Numerical studies have shown that very soon damping on small scales leads to a maximally
helical field (for which either $\A_+$ or $\A_-$ vanishes) which then continues to evolve
via an inverse cascade, see~\cite{Banerjee:2004df, Campanelli:2007tc}. The inverse cascade
is active as long as the Reynolds number of the cosmic fluid at the scale under consideration
is larger than one and the fluid is therefore turbulent~\cite{Biskamp:2003}.
The damping scale $k_{\rm diss}(\tconf)$ is the scale at which the Reynolds number becomes
of order unity. On scales smaller than $k_{\rm diss}(\tconf)$ the magnetic field and the
turbulent motion of the fluid are damped exponentially by viscosity.

In the following we investigate how the spectrum of helical magnetic fields evolves during
the turbulent epoch. We first discuss the evolution of the correlation scale of the magnetic
field and the duration of the turbulent phase, before computing the magnetic energy spectrum
at the end of the inverse cascade. Here we assume that the reheating epoch is relatively
short and ends at $\tconf_*$. This corresponds to the reheating temperature $T_*$, and
because of radiation domination, we can approximately use $\tconf/\tconf_*\simeq T_*/T$
during the turbulent phase.

The helical magnetic field from inflation always has a blue spectrum and is therefore
dominated by the largest wavenumber crossing the Hubble scale at the end of inflation
or, for simplicity, reheating
\bdm
  \widetilde\rho_B(\tconf_*)\ \simeq\ k_*^4\,\CS^2(\fN)\,,
\qquad \mbox { with } \ \ k_* \simeq \H_* \,.
\edm
Accordingly, the correlation scale (which is roughly given by the scale at which the
power spectrum peaks) is initially $k_c(\tconf_*)=k_*$. As an example we consider
$T_*=10^{14}\,\GeV$ and find
\bea
  k_c(\tconf_*)&\ =\ k_*\ =\ \H_*\ =\ H_*a_*
    \ =\ \sqrt{\frac{a_{SB}}{3}}g_0^{1/3}g_*^{1/6}\frac{T_0}{m_P}T_* &
  \nonumber \\
  &\ \simeq\ 10^{-17}\,\GeV \left(\frac{g_*}{200}\right)^{1/6}
    \left(\frac{T_*}{10^{14}\,\GeV}\right) &
  \nonumber \\
  &\ \simeq\ 2 \times 10^{21}\,\Mpci \left(\frac{g_*}{200}\right)^{1/6}
    \left(\frac{T_*}{10^{14}\,\GeV}\right) \,. &
\eea
Here $a_{SB}$ is the Stefan-Boltzmann constant, $a_{SB}=\pi^2/15$ in our units, $g_*$
and $g_0=2$ denote the number of relativistic degrees of freedom at $\tconf_*$ and
today, respectively, and $T_0=2.73\,{\rm K}$ is the present CMB temperature.

Let us assume that the inverse cascade starts at $\tconf_*$. In Ref.~\cite{Campanelli:2007tc}
it was found that during the inverse cascade of a maximally helical magnetic field the total
rescaled energy density scales like
\bdm
  \widetilde\rho_B(\tconf)\ \propto\ (\tconf/\tconf_*)^{-2/3}
\edm
and the comoving correlation scale evolves in the same way
\bdm
  k_c(\tconf)\ =\ k_c(\tconf_*)\, (\tconf/\tconf_*)^{-2/3}
\edm
such that the ratio $\widetilde\rho_B/k_c$ which is proportional to the rescaled
helicity density remains
constant. This continues until $\tconf_{\rm fin}$, the time when the damping scale has
grown up to the correlation scale, $k_{\rm diss}(\tconf_{\rm fin})= k_c(\tconf_{\rm fin})$.
After $\tconf_{\rm fin}$ the inverse cascade and turbulence cease and the magnetic field
evolves solely by flux conservation on large scales
\bdm
  \frac{d\widetilde\rho_B}{d\ln k}(k,\tconf)\ =\
  \frac{d\widetilde\rho_B}{d\ln k}(k,\tconf_{\rm fin})
  \qquad \mbox{ for }\ \ \tconf>\tconf_{\rm fin} \,,\ k< k_{\rm diss}(\tconf)
\edm
and viscosity damping on small scales, $k>k_{\rm diss}(\tconf)$. At the end of the inverse
cascade, the correlation scale of magnetic field spectrum has moved to
\bdm
  k_{\rm fin}\ \equiv\ k_c(\tconf_{\rm fin})\ =\ k_*\, (T_{\rm fin}/T_*)^{2/3}
\edm
and the total energy density is reduced by the same factor.

\begin{figure}[t]
\begin{center}
\includegraphics[width=0.65\textwidth]{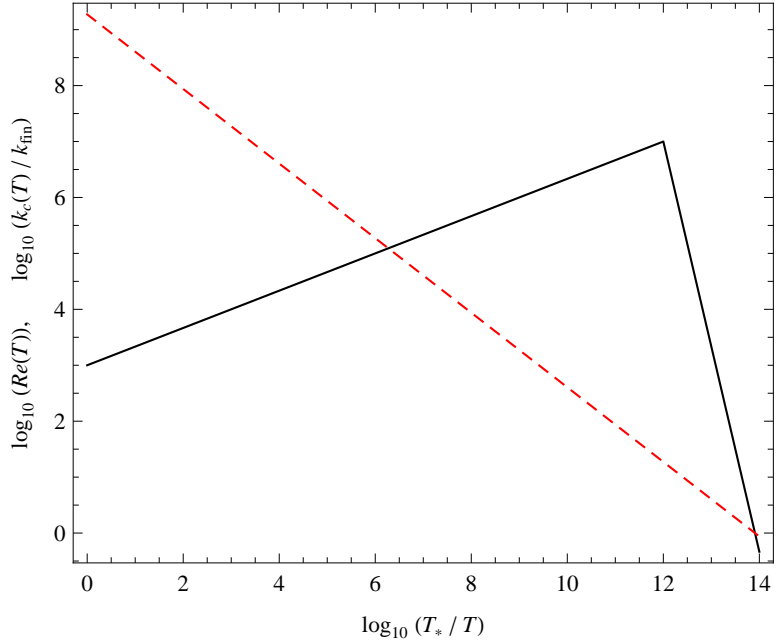}
\caption{The Reynolds number $\Rey(T)$ (black solid) and the comoving correlation scale
$k_c(T)/k_{\rm fin}$ (red dashed) are shown in log scale as a function of $\log(T_*/T)$ with
$T_*=10^{14}\,\GeV$. Turbulence and with it the inverse cascade terminate at
$T_{\rm fin}\simeq 1\,\GeV$ and $k_{\rm fin}=k_c(\tconf_{\rm fin}) \simeq (10^{-12}\,\Mpc)^{-1}$.}
\label{f:cascade}
\end{center}
\end{figure}

To compute the ratio $T_{\rm fin}/T_*$ we need to determine the temperature at which the
Reynolds number becomes unity. In Appendix~A of Ref.~\cite{Caprini:2009pr} the Reynolds
number at very high temperatures is estimated to be
\bdm
  \Rey(k,T)\ \propto\ \frac{aT}{k}\sqrt{\frac{\rho_B(k)}{\rho_f}}
\edm
for a given scale $k$. Here $\rho_f$ is the energy density of the fluid which contributes
to the turbulent motion. In perfect thermal equilibrium $\rho_f=\rho$.
More precisely, denoting the Reynolds number at the beginning of the inverse cascade by
$R_*=\Rey(k_c(\tconf_*),T_*)$, it is found
\be\label{e:ReT}
  \Rey(k_c(T),T)\ =\ \left\{ \begin{array}{ll}
    R_* \left(\frac{T_*}{T}\right)^{1/3} &\mbox{ for }\ T>T_{\rm ew}=100\,\GeV
    \vspace{2mm} \\
    R_* \left(\frac{T_*}{T_{\rm ew}}\right)^{1/3}
    \left(\frac{T}{ T_{\rm ew}}\right)^{11/3} & \mbox{ for }\ T<T_{\rm ew}\,.
  \end{array} \right.
\ee
Setting $g_*\simeq 200$ and $T_*=10^{14}\,\GeV$ yields
$R_* \simeq a_*T_*/k_c(\tconf_*) \simeq \sqrt{3/(a_{SB}g_*)}\,m_P/T_* \isorder{10^3}$.
(For simplicity we have set $\rho_B/\rho_f\sim 1$ for this value.) Note that after
inflation, until the electroweak transition at $T_{\rm ew}$, the Reynolds number of the fluid
is actually increasing. This comes from the fact that it is inversely proportional to the
comoving mean free path which is constant at early times. After the electroweak phase
transition, collisions are much more strongly suppressed and the comoving mean free path
grows like $a^4$, hence the Reynolds number decreases rapidly. For more details see
Ref.~\cite{Caprini:2009pr}. In \figref{f:cascade} we show the evolution of both, the Reynolds
number and the correlation scale of the helical magnetic field through the inverse
cascade.

Finally, we can now derive the generic scaling of $T_{\rm fin}$ with the initial temperature
$T_*$. With the scaling $k_{\rm fin} = k_*(T_{\rm fin}/T_*)^{2/3}$ and the help
of \eqref{e:ReT} for the evolution of the Reynolds number, we find that the Reynolds number
becomes unity and the inverse cascade stops at $T_{\rm fin}$ given by
\be
  \frac{T_{\rm fin}}{T_*}\ \simeq\ 10^{-14}\left(\frac{10^{14}\,\GeV}{T_*}\right)^{9/11}\,,
    \qquad \mbox { for } \ \ T_*>T_{\rm ew}
\ee
so that
\be\label{e:kfin}
  k_{\rm fin}\ \simeq\ 10^{12}\,\Mpc^{-1} \left(\frac{T_*}{10^{14}\,\GeV} \right)^{5/11} \,.
\ee
For $T_*=10^{14}\,\GeV$ we obtain $T_{\rm fin}\simeq 1\,\GeV$ and the correlation scale
moves by about 9 orders of magnitude from $k_*$ to $k_{\rm fin}$, see also \figref{f:cascade}.

\begin{figure}[t]
\begin{center}
\includegraphics[width=0.65\textwidth]{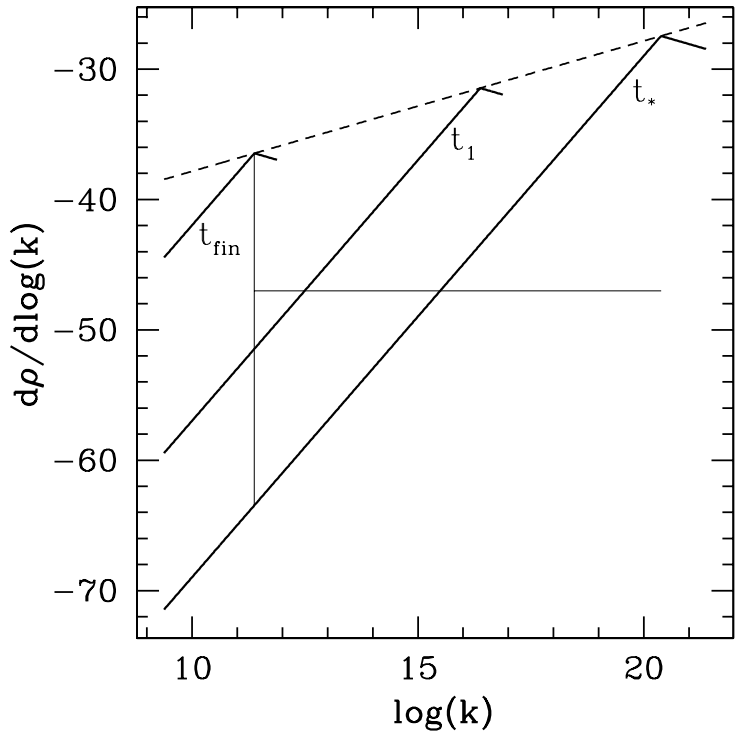}
\caption{The process of inverse cascade is depicted. We show
$\frac{d\widetilde\rho_B}{d\ln k}(k_c(\tconf),\tconf)$ as a function of $k_c(\tconf)$
(dashed line) as well as $\frac{d\widetilde\rho_B}{d\ln k}(k,\tconf_*)$,
$\frac{d\widetilde\rho_B}{d\ln k}(k,\tconf_1)$ and
$\frac{d\widetilde\rho_B}{d\ln k} (k,\tconf_{\rm fin})$ as functions of $k$ (thick solid
lines). We choose $\tconf_1/\tconf_*= 10^{6}$. The parts of the curves to the right of
the maximum are not reliable. The vertical line indicates the total amplification factor
which is constant for $k<k_0=k(\tconf_{\rm fin})$. The horizontal line indicates the amount
by which the correlation scale increases during the inverse cascade. The energy density is
in units of $\Gauss^2$ for $\CS(\fN)=1$ and $k$ is in units of $\Mpc^{-1}$.}
\label{f:casc2}
\end{center}
\end{figure}

Now we can trace the magnetic field spectrum through the inverse cascade. The spectral
shape on large scales remains unchanged~\cite{Campanelli:2007tc}.
At late time, $\tconf>\tconf_{\rm fin}$ we therefore obtain
\be\label{e:rhoBlate}
  \frac{\dd \widetilde\rho_B}{\dlnk}(\tconf,k)\ \simeq\ \left\{ \begin{array}{ll}
    \frac{\dd \widetilde\rho_B}{\dlnk}(\tconf_*,k_*)\, \left(\frac{k}{k_*}\right)^4
    \left(\frac{T_*}{T_{\rm fin}}\right)^2 & \mbox{ for }\ \ k<k_{\rm fin}
  \\   0 & \mbox{ else.}
\end{array} \right.
\ee
With $T_*/T_{\rm fin} \simeq 10^{14}$ this yields an amplification of the initial
amplitude by 28 orders of magnitude on scales larger than $1/k_{\rm fin}$, see the
sketch in \figref{f:casc2}.

For a generic spectral index $n$, \eqref{e:rhoBlate} is replaced by, see~\cite{Caprini:2009pr},
\be\label{e:rhoBlaten}\hspace{-2cm}
  \frac{\dd \widetilde\rho_B}{\dlnk}(\tconf,k)\ \simeq\ \left\{ \begin{array}{ll}
    \frac{\dd \widetilde\rho_B}{\dlnk}(\tconf_*,k_*)\left(\frac{k}{k_*}\right)^{n+3}
    \left(\frac{T_*}{T_{\rm fin}}\right)^{2(n+2)/3}
    & \mbox{ for }~ k<k_{\rm fin}\, , ~ \tconf>\tconf_{\rm fin} \\
  0 & \mbox{ else.}
\end{array} \right.
\ee
Clearly, the smaller $n$ the less significant is the amplification by the inverse cascade
and for $n=-2$ there is no amplification at all. For $n<-2$ the above result does not
apply, see~\cite{Caprini:2009pr}.

In our case, where $n=1$, \eqref{e:rhoM} can be written as
\bdm
  \frac{\dd \widetilde\rho_B}{\dlnk}(\tconf_*,k_*)
    \ \simeq\ 4 \times 10^{-68} \, \GeV^4\,\CS^2(\fN)
    \left(\frac{g_*}{200}\right)^{2/3} \left(\frac{T_*}{10^{14}\,\GeV}\right)^4
\edm
with which we arrive at
\bdm
  \frac{\dd \widetilde\rho_B}{\dlnk}(\tconf_{\rm fin},k_{\rm fin})
  \ \simeq\ 2 \times 10^{-77}\,\GeV^4\,\CS^2(\fN) \left(\frac{T_*}{10^{14}\,\GeV}\right)^{38/11}
\edm
where we assumed $g_*\simeq 200$. This determines the final strength of the magnetic field
on large scales\footnote{The magnetic field strength is $B=\sqrt{8\pi\rho_B}$ in Gauss
units, where $\GeV^2\simeq 1.4 \times 10^{19}\,\Gauss$, while in Heaviside-Lorentz units
we have $B=\sqrt{2\rho_B}$ and $\GeV^2/\sqrt{4\pi} \simeq 1.4 \times 10^{19}\,\Gauss$.},
\be \label{e:Bfinkfin}
 \widetilde B(k)\ \simeq\ 3 \times 10^{-19} \,\Gauss \ \CS(\fN)
   \left(\frac{k}{k_{\rm fin}}\right)^2 \left(\frac{T_*}{10^{14}\,\GeV}\right)^{19/11}
\ee
for $k\le k_{\rm fin}$. With the help of \eqref{e:kfin} this can also be written as
\be\label{e:Bfink}
  \widetilde B(k)\ \simeq\ 3 \times 10^{-19} \,\Gauss \ \CS(\fN)
    \left(\frac{k}{10^{12}/\Mpc}\right)^2 \left(\frac{T_*}{10^{14}\,\GeV}\right)^{9/11}
\ee
for $k\le k_{\rm fin}$.

After the end of the turbulent phase, magnetic fields are damped on small scales by
viscosity and evolve by flux conservation, so that $\widetilde B={\rm const.}$ on large
scales. For our typical value of $T_* \simeq  10^{14}\,\GeV$ hence
$k_{\rm fin}\simeq 10^{12}/\Mpc$, for cosmologically interesting scales of $k\sim 10/\Mpc$
the magnetic field is of the order of $\widetilde B(k=10/\Mpc)\simeq  10^{-40}\,\Gauss$.
This is much too small for dynamo amplification. For smaller reheating temperatures,
$T_*$, the Reynolds number grows less strongly and turbulence and the associated
inverse cascade are of shorter duration. Therefore the value of $\widetilde B(k)$ at
fixed $k<k_{\rm fin}$ is actually smaller for $T_*<10^{14}\,\GeV$ even though $k_*$ is
larger for a smaller reheating temperature, see \eqref{e:Bfink}. Considering
the lowest value for which our treatment
is valid, $T_* \simeq T_{\rm ew} \simeq 100\,\GeV$ we arrive at $T_{\rm fin}\simeq 8\,\MeV$
and $k_{\rm fin} \simeq 4 \times 10^6/\Mpc$ but the magnetic field is only
\be
  \widetilde B(k)\ \simeq\ 6 \times 10^{-40} \,\Gauss\ \CS(\fN)
  \left(\!\frac{k}{k_{\rm fin}}\!\right)^2 \,, \ \ \mbox{for}\ \ k\le k_{\rm fin} \simeq 10^{6}/\Mpc\,.
\ee

At scales of $0.1\,\Mpc$, this field is by far insufficient for subsequent dynamo
amplification which requires seed fields of the order of at least
$10^{-20}\,\Gauss$~\cite{Brandenburg:2004jv}. For an arbitrary reheating temperature
$T_*$ we obtain from~(\ref{e:Bfink})
\be \label{e:B0.1Mpc}
 \widetilde B(k=10/\Mpc)\ \simeq\ 3\times 10^{-41} \,\Gauss\ \CS(\fN)
   \left(\frac{T_*}{10^{14}\,\GeV}\right)^{9/11}
\ee
Let us investigate how large this field can become in the best case in which our
treatment may apply. Certainly we want to require backreaction to be unimportant, but we
do not insist that $f_N$ be small. Using (\ref{e:backT}) this requires
$\CS(\fN)<(m_P/T_*)^2$ and hence
\be \label{e:B0.1MpcT}
  \widetilde B(k=10/{\rm Mpc})\ \le\ 10^{-32}\,\Gauss
    \left(\frac{10^{14}\,\GeV}{T_*}\right)^{13/11}
\ee
To achieve a minimal necessary field for dynamo amplification of $B\sim 10^{-20}\,\Gauss$ we would need
$T_* \le 10^4\,\GeV$, a somewhat low inflation scale, but not excluded.

Until recombination, fields on scales smaller than about 
$0.1(B/10^{-9}\Gauss)\,\Mpc$ are damped by viscosity~\cite{PhysRevD.57.3264}. 
The scales $k\gsim 10(10^{-9}\Gauss/B)/\Mpc$ therefore do not survive the 
linear regime and will not be amplified before being damped. But even on 
these smallest ``surviving scales''
the magnetic field generated is in most cases too weak for dynamo amplification.

%%%%%%%%%%%%%%%%%%%%%%%%%%%%%%%%%%%%%%%%%%%%%%%%%%%%%%%%%%%%%%%%%%%%%%%%%
\section{Conclusions} \label{sec:conclusions}
%%%%%%%%%%%%%%%%%%%%%%%%%%%%%%%%%%%%%%%%%%%%%%%%%%%%%%%%%%%%%%%%%%%%%%%%%

In this work, we have studied the generation of helical magnetic fields during inflation
by adding a parity-violating term of the form $f(\phi)F\Fdual$ to the action. For
two specific choices
of the coupling function $f(\phi)$, namely a power law and an exponential form,
we found that, during slow roll inflation, the power spectrum of the generated magnetic
fields is always blue with spectral index $n=1$.
The aspect  responsible for this result is that the helical coupling term in the
evolution equation for the Fourier modes of the vector potential is active only around
horizon crossing of a given mode and it is not varying rapidly during slow roll inflation.
As a consequence, in slow roll inflation all Fourier modes are amplified in the same way 
and the final shape of the power spectrum is unchanged with respect to the initial one.

We have also studied the effects of a short deviation from slow roll.
If kept within the bounds allowed by the CMB data, such a departure from slow roll
does not strongly
modify the overall shape of the magnetic field power spectrum. In principle, a very
steep coupling function, e.g.~a double-exponential $f(\phi)=f_0\exp[\exp(\al\phi/m_P)]$,
can lead to a different spectrum, $n\neq 1$. However, if we want to satisfy
perturbativity, $|\fN|< 1$,
for such a behaviour to be valid throughout inflation, a ridiculously small pre-factor
$f_0$ is required. In the absence of a convincing physical motivation, we have 
 not investigated such an extreme case.

After inflation and reheating, an inverse cascade sets in and moves power from the
correlation scale $k_*\simeq \H_*$ to larger scales, which is schematically represented
in \figref{f:casc2}. We find that typical values for the reheating temperature,
e.g.~$T_*\simeq 10^{14}\,\GeV$, can only lead to magnetic fields of the order of
$10^{-40}\,\Gauss$ on scales of 0.1 Mpc and smaller values are obtained for smaller
reheating temperatures.  {Unless $\CS(f_N)\gg 1$ and the reheating temperature is 
sufficiently low, $T_* \le 10^4$GeV, }  these field amplitudes are largely insufficient 
for dynamo amplification~\cite{Brandenburg:2004jv}.

This result is quite generic. A power law or an exponential coupling to the
$F\Fdual$ term during slow roll inflation will generate sufficiently strong seeds
for the large scale magnetic fields in galaxies and clusters without also
leading to backreaction only if the inflation scale is sufficiently low, of the order
of the electroweak scale.  Even though for helical fields an inverse cascade
does move power to larger scales, this is still insufficient in most cases.
For typical inflation scales, to obtain sufficient
fields on large scales after the inverse cascade, they must have been so large on small
scales after inflation that backreaction cannot be neglected. Another possibility might
be a very steep coupling which can lead to a different, $n\neq 1$ spectrum. However,
only if the spectrum is close to scale invariant, it can have significant amplitudes
on large scales without significant backreaction from small scales.

Scenarios of inflationary magnetogenesis are often constrained by requiring that the
backreaction of the generated magnetic field on the background evolution is small. Since
the perturbations in the scalar field are affected by the presence of a non-minimal coupling
as indicated in \eqref{e:dephi}, it will be interesting to study the backreaction effects
of the EM perturbations on the evolution of the primordial curvature perturbations. This may
provide another tool to constrain inflationary scenarios of magnetogenesis. This issue will
be addressed in a future project~\cite{WeInPrep}.

% --- ACKNOWLEDGMENTS --- %%%%%%%%%%%%%%%%%%%%%%%%%%%%%%%%%%%%%%%%%%%%%%
\section*{Acknowledgments}
We would like to thank Mohamed Anber and Lorenzo Sorbo for discussions.
We specially thank Claudia de Rham for some very fruitful discussions.
We would also like to thank the anonymous referee for his/her constructive suggestions
which helped in improving the content of the paper.
The authors acknowledge financial support from the Swiss National Science Foundation.
R.K.J.~acknowledges support from a research fellowship of the Indo Swiss Joint Research
Program with grant no. RF~12.

% --- APPENDICES --- %%%%%%%%%%%%%%%%%%%%%%%%%%%%%%%%%%%%%%%%%%%%%%%%%%%%
\appendix

%%%%%%%%%%%%%%%%%%%%%%%%%%%%%%%%%%%%%%%%%%%%%%%%%%%%%%%%%%%%%%%%%%%%%%%%%
\section{Quantisation of the vector potential} \label{app:quantisation}
%%%%%%%%%%%%%%%%%%%%%%%%%%%%%%%%%%%%%%%%%%%%%%%%%%%%%%%%%%%%%%%%%%%%%%%%%

Always working in Coulomb gauge, we promote the vector potential $A_i$ to a quantum
mechanical operator, define conjugate momentum as $\Pi^i\equiv\delta S/\delta\dot{A}_i$,
and impose the commutation relation
\be
  \left[ A_i(\tconf,\Vx),\ \Pi^j(\tconf,\Vy) \right]\ =\ i \delta_{\bot i}^j(\Vx-\Vy) \,.
  \label{eq:APi_comm}
\ee
Here the transversal Dirac delta, $\delta_{\bot i}^j$, is defined as
\be
  \delta_{\bot i}^j(\Vx-\Vy)\ \equiv\ \int\frac{\dkkk}{(2\pi)^3} e^{i\Vk\cdot(\Vx-\Vy)}
    \left(\delta_i^j-\hat{k}_i\hat{k}_m\delta^{mj}\right)
\ee
with $\hat{k}_i\equiv k_i/k$ and $k\equiv |\Vk|$. Let us compute the conjugate momentum
in terms of the vector potential which we expand in terms of creation and annihilation
operators in Fourier space. The canonical commutation relations of the creation and
annihilation operators, together with the above commutation relation of $A_i$ and $\Pi^j$
will lead to a Wronskian normalisation condition on the mode functions of the quantised
vector potential.

The conjugate momenta in the perturbed FL metric are
\be
  \Pi^i\ =\ a^2 g^{i j}\left(\dot{A}_j +f(\varphi)\epsilon_{jmn}\partial_m A_n\right)\,.
  \label{eq:conjmom}
\ee
The non-standard second term arises due to the axial coupling of the EM field to the
inflaton. Since it involves the curl of the vector potential it is convenient to expand
the operators in Fourier space with respect to a helicity basis. For each comoving wave
vector, $\Vk$, we define the comoving right-handed orthonormal basis
\be
  e^0_\mu\ \equiv a \Big(1,\,\V{0}\Big) \,, \quad
  e^{1,2}_\mu(\Vk)\ \equiv\ a \Big(0,\,\Veps^\Vk_{1,2}\Big) \,, \quad
  e^3_\mu(\Vk)\ \equiv a \Big(0,\,\Vkhat\Big)
\ee
with
\be
  \Vkhat\cdot\Veps^\Vk_{1,2}\ =\ 0\ =\ \Veps^\Vk_1\cdot\Veps^\Vk_2 \,,
  \qquad \Veps^\Vk_1\wedge\Veps^\Vk_2\ =\ \Vkhat \,. \quad |\Veps^\Vk_{i}|^2=1
\ee
and
\be
  \Veps_1^{-\Vk}\ =\ -\Veps_1^\Vk \,, \qquad \Veps_2^{-\Vk}\ =\ \Veps_2^\Vk \,.
\ee
The two transverse directions are combined to the helicity (or circular) directions
\be
  \Veps^\Vk_\pm\ \equiv\ \frac{1}{\sqrt{2}}\left( \Veps^\Vk_1 \pm i\Veps^\Vk_2
   \right) \,.
\ee
These have the following useful properties
\bea
  \Veps^{\Vk*}_\pm\ =\ \Veps^\Vk_\mp
  \\
  \Veps^{-\Vk}_\pm\ =\ -\Veps^\Vk_\mp
  \\
  i\Vkhat\wedge\Veps^\Vk_\pm\ =\ \pm\Veps^\Vk_\pm
  \\
  \Veps^\Vk_h \cdot \Veps^{\Vk*}_{h'}\ =\ \delta_{hh'}
  \\
  \sum_h \varepsilon^\Vk_{h,i} \, \varepsilon ^\Vk_{h,j}
    \ =\ \delta_{ij} - \hat{k}_i \hat{k}_j
\eea
where $h,h'\in\{+,-\}$ and the star denotes complex conjugation. We now expand the
vector potential in Fourier space in the helicity basis.
\bea
  &A_j(t,\Vx)\ =\ \int\frac{\dkkk}{(2\pi)^3} \sum_{h=\pm} \Big\{ &
    e^h_j(\Vk) b_h(\Vk) A_h(t,k) e^{i\Vk\cdot\Vx}
  \nonumber \\
  && + e^{h*}_j(\Vk) b^\dagger_h(\Vk) A^*_h(t,k) e^{-i\Vk\cdot\Vx} \Big\}
\eea
where the creation and annihilation operators, $b^\dagger_h(\Vk)$ and
$b_h(\Vk)$, satisfy the canonical commutation relations
\bea
  \left[ b_h(\Vk),\, b_{h'}^\dagger(\Vq) \right]\ =\ (2\pi)^3\, \delta^3(\Vk-\Vq)\, \delta_{hh'}
  \\
  \left[ b_h(\Vk),\, b_{h'}(\Vq)\right]\ =\ 0\ =\   \left[ b^\dagger_h(\Vk),\, b^\dagger_{h'}(\Vq)\right] \,.
\eea
The vacuum, $|0\rangle$, is defined by $b_h(\Vk)|0\rangle\equiv 0$. As a consequence of
the commutation relation~(\ref{eq:APi_comm}) of $A_i$ and its conjugate momentum, $\Pi^j$,
as given in~(\ref{eq:conjmom}), the mode function must satisfy the Wronskian normalisation
condition. In terms of the rescaled Fourier modes of the vector potential,
$\A_h(t,k)\equiv a A_h(t,k)$, this reads
\be   \label{eq:wronskian}
  \dot\A_h(\tconf,k)\A_h^*(\tconf,k) - \A_h^*(\tconf,k)\dot\A_h(\tconf,k)\ =\ i \,.
\ee
Finally, the evolution equation for the Fourier modes can now be derived from the forced
wave equation of the vector potential, \eqref{eq:lin_maxinhom},
\be
  \ddot\A_h(\tconf,k) +\left[k^2+hkf'(\varphi)\dot\varphi\right]\A_h(\tconf,k)\ =\ 0
\ee
where $h=\pm$ reflects the two helicity modes.

%%%%%%%%%%%%%%%%%%%%%%%%%%%%%%%%%%%%%%%%%%%%%%%%%%%%%%%%%%%%%%%%%%%%%%%%%
\section{Perturbative chiral interaction}
\label{app:perturbativity}
%%%%%%%%%%%%%%%%%%%%%%%%%%%%%%%%%%%%%%%%%%%%%%%%%%%%%%%%%%%%%%%%%%%%%%%%%

In order to derive a bound on the parameter space of the coupling function $f(\phi)$, be it a power law, an exponential or any other functional form, we demand the interaction between the scalar field and electromagnetism to be small in order for perturbative quantum field theory to make sense. For example, we want the interaction picture to apply and the first terms in the Feynman-Dyson series for the scattering matrix to be small. Therefore we demand that $|S_{I}[\phi,A_\mu]|<|S_{\rm em}[A_\mu]|$.

Let us express the free EM Lagrangian density in terms of the electric and magnetic energy densities (as defined in \secref{sec:em})
\be
  \mathcal{L}_{\rm em}\ =\ -\frac{1}{4}F^2\ =\ \frac{1}{2}\left(E^2-B^2\right)
  \ =\ \rho_E-\rho_B
\ee
and compute the free EM action in the FL background:
\bea
  S_{\rm em}[A_\mu] &=& \int\dd^4x \sqrt{-g}\, \mathcal{L}_{\rm em}
  \ =\ \int\dd^4x\, a^4 \left(\rho_E-\rho_B\right)
  \nonumber \\
  &=& \int\dtconf \int\dd^3x\, \left(\widetilde{\rho}_E-\widetilde{\rho}_B\right)
  \ =\ V \int\dtconf\, \left\langle\widetilde{\rho}_E(\tconf)-\widetilde{\rho}_B(\tconf)\right\rangle_{V}
\eea
where we rescaled the energy densities as $\widetilde\rho_B\equiv a^4\rho_B$ and defined 
the volume average $\langle\ldots\rangle_{V}$ in the last step. We assume that for 
quantum fluctuations the volume average equals the vacuum expectation value, 
$\langle\ldots\rangle$, such that
\bea
  \langle\widetilde\rho_B(t)\rangle_{V} &=& \frac{1}{2}\,\langle\widetilde{B}_i\widetilde{B}^*_j\rangle\delta^{ij}
  \ =\ \int_0^\infty\frac{k^4\dk}{(2\pi)^2}\Big[ |\A_+(\tconf,k)|^2 + |\A_-(\tconf,k)|^2 \Big]
  \\
  \langle\widetilde\rho_E(t)\rangle_{V} &=& \frac{1}{2}\,\langle\widetilde{E}_i\widetilde{E}^*_j\rangle\delta^{ij}
  \ =\ \int_0^\infty\frac{k^2\dk}{(2\pi)^2}\Big[ |\dot\A_+(\tconf,k)|^2 + |\dot\A_-(\tconf,k)|^2 \Big]\,,
\eea
as we show in \secref{sec:ps}.

To be able to compute the contribution of the interaction $f(\phi)F\Fdual$ to the action we pass to the language of differential forms by writing the electromagnetic 2-form
$F=\frac{1}{2}F_{\mu\nu}dx^\mu dx^\nu =dA = d(A_\mu dx^\mu)$, such that
\be
  F\Fdual \sqrt{-g}\, \dd^4x \ =\ 2 F \wedge F \,.
\ee
Then we use the facts that $F=dA$ and $dF=d(dA)=0$ to integrate by parts
\bea
  S_{I}[\phi,A_\mu] &=& \frac{1}{4}\int \dd^4x \sqrt{-g}\, f(\phi) F\Fdual
  \ =\ \frac{1}{2}\int f(\phi) dA \wedge F
  \nonumber \\
  &=& \frac{1}{2}\int \Big( d\big[f A \wedge F\big] - df \wedge A \wedge F \Big)
  \ =\ \frac{1}{2}\int df \wedge F \wedge A \,.
\eea
In the last step we assume the boundary terms to vanish. Clearly, $S_{I}[\phi,A_\mu]$ vanishes for a constant $f(\phi)$. Next, we express $F$ in terms of $E$ and $B$ as measured by an observer with timelike four-velocity $u$, see \secref{sec:em}:
\be
  F\ =\ u\wedge E - 2\widetilde{(u\wedge B)} \,.
\ee
Using this we find
\be
  df \wedge F \wedge A\ =\ df \wedge u \wedge E \wedge A
  + 2f_{,\alpha} u^{[\alpha} B^{\beta]} A_\beta\, \sqrt{-g}\, \dd^4x \,.
\ee
In a FL background, the only valid observer is the Hubble flow, $u=-a\dtconf$, and $f$ is only a function of time, i.e.~$df=\dot{f}\dtconf$, such that $df\wedge u=0$, $f_{,\alpha}u^\alpha=a^{-1}\dot{f}$ and $f_{,\alpha}B^\alpha=0$. So we find
\be
  df \wedge F \wedge A\ =\ \dot{f}\, B^\beta A_\beta\, a^3 \, \dd^4x
  \ =\ \dot{f}\, \widetilde{\hel} \, \dd^4x
\ee
where $\widetilde{\hel}$ is the rescaled helicity density which we introduce in \secref{sec:em}. Finally, we can integrate to find the action
\be
  S_{I}[\phi,A_\mu]\ =\ \frac{1}{2}\int \dd^4x\, \dot{f}\, \widetilde{\hel}
  \ =\ \frac{1}{2}V \int\dtconf\, \dot{f}\, \langle\widetilde{\hel}(\tconf)\rangle_{V} \,.
\ee
And again, the volume average is given by the expectation value
\be
  \langle\widetilde{\hel}(\tconf)\rangle_{V}
  \ =\ \langle\widetilde{B}_i\widetilde{A}^*_j\rangle\delta^{ij}
  \ =\ 2 \int_0^\infty\frac{k^3\dk}{(2\pi)^2}
    \Big[ |\A_+(\tconf,k)|^2 - |\A_-(\tconf,k)|^2 \Big]\,.
\ee
using the results given in \secref{sec:ps}. 

The implications for the coupling function from the condition $|S_{I}[\phi,A_\mu]|
< |S_{\rm em}[A_\mu]|$ using the results derived here are discussed 
in~\secref{sec:coupf}.

%%%%%%%%%%%%%%%%%%%%%%%%%%%%%%%%%%%%%%%%%%%%%%%%%%%%%%%%%%%%%%%%%%%%%%%%%
\section{Asymptotic behaviour of the Coulomb wave functions}\label{app:coulomb}
%%%%%%%%%%%%%%%%%%%%%%%%%%%%%%%%%%%%%%%%%%%%%%%%%%%%%%%%%%%%%%%%%%%%%%%%%

To investigate the late time behaviour of the solution to the mode equation for a
power law coupling, we need asymptotic expressions for the Coulomb wave function,
$G_0(y,x)$ and $F_0(y,x)$, for $x\to 0$. Based on their expansion in terms of Bessel
functions given in Abramowitz \& Stegun~\cite{Abramowitz:1972}, we derive the
asymptotes of $G_0$, $F_0$, $G'_0$, $F'_0$ for $x\to 0$ for arbitrary but fixed $y$.
Actually, we argue that the asymptotic limits for $x\ll2y$ given in 14.6.9-10 (p.~542)
of~\cite{Abramowitz:1972} are incorrect. All statements given here were verified
numerically using \textit{Mathematica} \cite{Mathematica7} and \textit{Maple} \cite{Maple13}.

We start from the asymptotic expressions in terms of the modified Bessel functions, $K$
and $I$, given in 14.6.7 of~\cite{Abramowitz:1972} for $L=0$, and $2y\gg x$
\bea
  G_0(y,x)\ \simeq\ 2\sqrt{2xy}\, W(y) K_1(2\sqrt{2xy})
  \\
  F_0(y,x)\ \simeq\ \frac{x}{\sqrt{2xy}} W(y)^{-1}\, I_1(2\sqrt{2xy})
\eea
where (see $C_0$ given in 14.1.8 of~\cite{Abramowitz:1972})
\be
  W(y)\ \equiv\ \left[ \frac{e^{\pi y}\sinh(\pi y)}{\pi y} \right]^{1/2} \,.
\ee
We notice that the approximations given in 14.6.8 of A\&S, where
\be
  \sinh \pi y\ \simeq\ \frac{1}{2} e^{\pi y}
\ee
is used, are only accurate to better than 1\% if $y\gtrsim 1$. Since in our case $y$ is
arbitrary or rather smaller than unity, we cannot employ this approximation. As for
the $x$-derivatives of $G_0$ and $F_0$ we have verified numerically that the derivatives
of the above expressions are good approximations,
\bea
  G'_0(y,x)\ \simeq\ -4y W(y)\, K_0(2\sqrt{2xy})
  \\
  F'_0(y,x)\ \simeq\ W(y)^{-1}\, I_0(2\sqrt{2xy}) \,.
\eea

Next we use the asymptotic properties of the modified Bessel functions for \emph{small}
arguments. As given on p.~375 in~\cite{Abramowitz:1972}
\bea
  & K_1(z)\ \simeq\ 1/z +\order{z} \,,\qquad
  & K_0(z)\ \simeq\ -\ln(z/2) -\gamma_E +\order{z^2}
  \\
  & I_1(z)\ \simeq\ z/2 +\order{z^3} \,,\qquad
  & I_0(z)\ \simeq\ 1 +\order{z^2}
\eea
for $z=2\sqrt{2xy}\ll1$. Using these expressions we find to lowest order in $x$ for
$x\ll (8y)^{-1}$
\bea
  & G_0(y,x)\ \simeq\ W(y) \,,\qquad
  & G'_0(y,x)\ \simeq\ 2yW(y)\ln(2xy)
  \\
  & F_0(y,x)\ \simeq\ x/W(y) \,,\qquad
  & F'_0(y,x)\ \simeq\ 1/W(y) \,.
\eea

These asymptotic expressions differ significantly from those given in 14.6.9-10
of~\cite{Abramowitz:1972} and it is straightforward to understand why: one arrives at
the expressions of~\cite{Abramowitz:1972} when falsely using the asymptotic expansions
of the modified Bessel functions for \emph{large} arguments, where
$I_1(|z|\gg 1)\simeq e^z/\sqrt{2\pi z}$ and $K_1(|z|\gg 1)\simeq e^{-z}\sqrt{\pi/(2z)}$.
However, for $x\to 0$ and fixed $y$, $z=2\sqrt{2xy}$ tends to zero and is not large as
assumed for approximations 14.6.9-10 of~\cite{Abramowitz:1972}.

% --- REFERENCES --- %%%%%%%%%%%%%%%%%%%%%%%%%%%%%%%%%%%%%%%%%%%%%%%%%%%%
\section*{References}
\bibliography{hmf_refs}
\bibliographystyle{JHEP}

%%%%%%%%%%%%%%%%%%%%%%%%%%%%%%%%%%%%%%%%%%%%%%%%%%%%%%%%%%%%%%%%%%%%%%%%%%%%%%%%%%%%%
\end{document}